 \documentclass[namedreferences]{solarphysics} 
\usepackage[optionalrh,solaromanenum]{spr-sola-addons} 
\usepackage[]{graphicx}                    
\usepackage{amssymb}                    
\usepackage{color}                       
\usepackage{url}                         

\usepackage{epstopdf}
\usepackage{ulem} 

\graphicspath{{./}} 

\begin{document}

\begin{article}

\begin{opening}

\title{PICARD SODISM, a space telescope to study the Sun from the middle ultraviolet to the near infrared
}


\author{M.~\surname{Meftah}$^{1}$\sep
        J.-F.~\surname{Hochedez}$^{1,7}$\sep
        A.~\surname{Irbah}$^{1}$\sep
        A.~\surname{Hauchecorne}$^{1}$\sep
        P.~\surname{Boumier}$^{2}$\sep
        T.~\surname{Corbard}$^{3}$\sep
        S.~\surname{Turck-Chi\`eze}$^{4}$\sep
        P.~\surname{Assus}$^{3}$\sep
        E.~\surname{Bertran}$^{1}$\sep
        P.~\surname{Bourget}$^{8}$\sep
        F.~\surname{Buisson}$^{5}$\sep
        M.~\surname{Chaigneau}$^{2}$\sep
        L.~\surname{Dam\'e}$^{1}$\sep
        D.~\surname{Djafer}$^{6}$\sep
        C.~\surname{Dufour}$^{1}$\sep
        P.~\surname{Etcheto}$^{5}$\sep
        P.~\surname{Ferrero}$^{1}$\sep
        M.~\surname{Hers\'e}$^{1}$\sep
        J.-P.~\surname{Marcovici}$^{1}$\sep
        M.~\surname{Meissonnier}$^{1}$\sep
        F.~\surname{Morand}$^{3}$\sep
        G.~\surname{Poiet}$^{1}$\sep
        J.-Y.~\surname{Prado}$^{5}$\sep
        C.~\surname{Renaud}$^{3}$\sep
        N.~\surname{Rouanet}$^{1}$\sep
        M.~\surname{Rouz\'e}$^{5}$\sep
        D.~\surname{Salabert}$^{3}$\sep
        A.-J.~\surname{Vieau}$^{1}$
       }
\runningauthor{Meftah et al.}
\runningtitle{SODISM, an imaging space telescope to observe the Sun from the MUV to the NIR}

\institute{ $^{1}$ LATMOS - {\it Laboratoire Atmosph\`eres, Milieux, Observations Spatiales}, CNRS -- Universit\'e Paris~VI \& Universit\'e de Versailles Saint-Quentin-en-Yvelines -- IPSL, F-78280, Guyancourt, France
\\Email: \url{Mustapha.Meftah@latmos.ipsl.fr} \\
            $^{2}$ IAS - {\it Institut d'Astrophysique Spatiale}, CNRS -- Universit\'e Paris~XI, F-91471 Orsay, France \\
            $^{3}$ OCA - {\it Observatoire de la C\^ote d'Azur}, Laboratoire Lagrange, Universit\'e de Nice-Sophia Antipolis, CNRS, Parc Valrose, F-06108 Nice Cedex 2 France \\
            $^{4}$ IRFU, CEA-Saclay, F-91191 Gif-sur-Yvette Cedex, France \\
            $^{5}$ CNES - {\it Centre National d'Etudes Spatiales}, Edouard Belin, F-31000 Toulouse, France \\
            $^{6}$ Unit\'e de Recherche Appliqu\'ee en Energies Renouvelables URAER/CDER, B.P. 88, Gharda\" ia, Algeria \\
            $^{7}$ ORB - {\it Observatoire Royal de Belgique}, All\'ee Circulaire 3., B-1180 Bruxelles (Uccle), Belgique\\
           Email: \url{Hochedez@oma.be} \\
            $^{8}$ ESO - {\it European Southern Observatory}, Chile\\
          }

\begin{abstract}
The {\it Solar Diameter Imager and Surface Mapper} (SODISM) onboard the PICARD space mission provides wide-field images of the photosphere and chromosphere of the Sun in five narrow pass bands (centered at 215.0, 393.37, 535.7, 607.1, and 782.2\,nm).
PICARD is a space mission, which was successfully launched on 15~June 2010 into a Sun synchronous dawn-dusk orbit.
It represents a European asset aiming at collecting solar observations that can serve to estimate some of the inputs to Earth climate models.
The scientific payload consists of the SODISM imager and of two radiometers, SOVAP (SOlar VAriability PICARD) and PREMOS (PREcision MOnitor Sensor), 
which carry out measurements that allow estimating the Total Solar Irradiance (TSI) and the Solar Spectral Irradiance (SSI) from the middle ultraviolet to the red.
The SODISM telescope monitors solar activity continuously.
It thus produces images that can also feed SSI reconstruction models.
Further, the objectives of SODISM encompass the probing of the interior of the Sun {\it via} helioseismic analysis of observations in intensity (on the solar disc and at the limb), and {\it via} astrometric investigations at the limb.
The latter addresses especially the spectral dependence of the radial limb shape, and the temporal evolution of the solar diameter and asphericity.
After a brief review of its original science objectives, this paper presents the detailed design of the SODISM instrument, its expected performance, and the scheme of its flight operations.
Some observations with SODISM are presented and discussed.

\end{abstract}
\end{opening}



\section{Introduction} \label{S-Introduction}

Even though most electromagnetic radiations in the spectral windows that range from the middle ultraviolet (MUV) to the near infrared (NIR) reaches the ground, there are compelling reasons to observe the Sun in these wavelengths from space instead \cite[and references therein]{Veselovsky2009ASR}. The rationale includes the prospects of: ``24/7/365'' temporal coverage, more accurate photometry, and seeing conditions unaltered by the Earth atmosphere.

Such factors have prevailed throughout the definition of the PICARD space mission, which is to expand observations of the global parameters of the Sun and primarily, to link the variability of its total and spectral irradiance with the presumptive changes of its geometric dimensions (chiefly of its diameter and oblateness), and with the fluctuations of solar magnetic activity \cite{Dame1999AdSpR,Thuillier2006ASR}.

Bringing consistency between these global solar parameters and the outcome of helioseismology, as well as mutual consistency among the former, has for long been expected to enable insights on the stellar interior \cite{Gilliland1982ApJ,Delache1993ApJ,Lefebvre2005MmSAI,Sofia2005MmSAI,Steiner2005AN,Stothers2006ApJ}. Beyond mass, age, elemental composition, or neutrino flux, the global parameters classically involve:
\begin{itemize}
\item The solar luminosity, and its spectral and temporal dependences,
\item The angular rotation velocity, and its radial, latitudinal and temporal dependences,
\item The center-to-limb variation (CLV), its outer part, {\it viz.} the limb darkening function (LDF), and their latitudinal and temporal dependences,
\item The photospheric radius and its dependence on latitude ({\it viz.} asphericity) and time,
\item The photospheric effective temperature, and its latitudinal and temporal dependences.
\end{itemize}

SODISM has been devised to address all five items above, and primarily the third and fourth ones, namely, estimating the LDF, the photospheric radius and the oblateness, in absolute terms (in view of their Schwabe cycle and secular evolutions beyond the few-year lifetime of the space mission), as well as in relative terms (by the measurement of their variations on the scale of months, if any). Nevertheless, in complementarity to the exploitation of dedicated radiometers such as PICARD-SOVAP \cite{Conscience2011SPIE} and PICARD-PREMOS \cite{Schmutz2009Metro}, SODISM investigations can also contribute to assess the SSI {\it via e.g.} the models that assign a known spectral radiance to particular solar features \cite{Fontenla2011JGRD}. In addition, SODISM data could be exploited in view of solar radiance studies, regarding {\it e.g.} its uniformity \cite{Rast2008ApJ}, or the dependence of the contrast of solar features on the center-to-limb location or on the cycle \cite{Crane2004A&A,Penn2007ApJ,Wesolowski2008SoPh}. Furthermore, sequences of photospheric and chromospheric images offer the possibility to determine the velocity field of the solar rotation and meridional flow, or characteristics of the supergranular pattern, which can be viewed as a global solar parameter too \cite{Meunier2008A&A}.

By aiming at filling the observational gap that concerns astrometry at the solar limb, and by therefore carrying out resolved observations in narrow pass bands free of Fraunhofer lines in the visible spectral range, fundamental features of its instrumental design qualify SODISM to monitor solar oscillations in intensity \cite{Corbard2008AN}. The inclusion of helioseismology among its objectives is scientifically sound and technically straightforward albeit increasing the telemetry requirement and the duty cycle of the instrument ({\it e.g.} use of the shutter). The other solar missions such as SDO-HMI (Solar Dynamics Observatory - Helioseismic and Magnetic Imager) aim at monitoring oscillations in velocity, for which the signal-to-noise is higher, but the physical complementarity of intensity and velocity measurements provides a strong motivation to exploit SODISM for helioseismology. The evidenced amplification of p modes at the limb \cite{Appourchaux1998IAU,Toner1999ApJ} constitutes a second incentive. This gain could indeed help detecting low degree low frequency pressure modes of oscillations (p~modes) and gravity modes of oscillations (g~modes) despite the fact that their detection is known to be challenging.

Let us recall that the photosphere represents a crucial region of the solar atmosphere. It is the deepest layer that is accessible to direct observation, hence the closest to the origins of magnetic activity. The photosphere is also the relatively small volume of space where the vast majority of the energy produced in the nuclear core of the Sun is radiated away, and where a minute fraction of it exudes to the corona by means that are still a matter of research.

As a result, the PICARD mission has been proposed to not only address the climatic parameters that emanate preponderantly from the photosphere and chromosphere, {\it i.e.} the total solar irradiance and the solar spectral irradiance \cite[and references therein]{Ermolli2012ACPD}, but to also contribute to the understanding of the other solar factors that might affect Earth and its climate, namely:
\begin{itemize}
\item The open magnetic field in the heliosphere, which shields the planets against the GCRs (galactic cosmic rays) and modulates their geoeffective flux,
\item The closed magnetic field and coronal activity that can produce SEPs (solar energetic particles) at the occasion of solar eruptions (flares, coronal mass ejections (CMEs), {\it etc.}),
\item The stationary solar wind, and its intermittent component, the CMEs.
\end{itemize}
The above causal factors are not directly observable by the PICARD payload, but the general strategy consists in developing a comprehensive knowledge that is based on models that conciliate all global parameters and that would permit to extrapolate the relevant solar agents (spectral irradiance, global magnetism, coronal activity, solar wind, {\it etc.}) toward the past, and eventually toward the future \cite{TurckChieze2007AdSpR}. The above line appears challenging but is there an alternative?

The PICARD program owes its name to Jean Picard (1620-1682), considered as a pioneer of precise modern astrometry who observed the sunspots and measured their rotation velocity and the solar diameter. The eponymous space mission has been promoted and funded by the French space agency (CNES), and the SODISM telescope has been designed and built between 1999 and 2010, under the responsibility of CNRS-LATMOS (previously `Service d'A\'eronomie').

The PICARD payload uses the MYRIADE family platform developed by CNES and designed for a total mass of about 130 kg at launch.
The spacecraft was successfully launched into a Sun-synchronous dawn-dusk orbit on 15~June 2010 by a DNEPR-1 launcher.
The mission was commissioned in-flight in October of the same year.

In the next section, we list and justify the design choices of the SODISM instrument, and in the following sections, we present its optical, imaging, mechanical, thermal, and operational details.


\section{Design Choices}\label{S-Choices}

As mentioned above, the Earth atmosphere generates various hindrances that make morphometry and photometry difficult. They include refraction and turbulence, not to speak of scattering, extinction, and of the diurnal alternation. It is suspected that the past inconsistencies regarding the temporal dependence of the solar diameter measured from the ground would stem primarily from such contingencies \cite{Ribes1991,Delache1994,Damiani2007MNRAS}. Avoiding them has been the major motivation for developing SODISM as a space imaging telescope. Learning if and how Earth atmospheric perturbations could be corrected for is the objective of a parallel project of the PICARD program, which acquires solar images from ground in the same channels as SODISM with a twin of it (the Qualification Model) installed in Calern located in the South of the French Alps \cite{Meftah2012SPIE,Irbah2011SPIE}.

Nevertheless, going to space does not resolve all issues, and it actually creates new ones. For SODISM to be a wide-field telescope while remaining compact and minimizing the different possible optical aberrations, a Ritchey-Chr\'etien design has been selected \cite{Wilson2007}. It consists of a concave hyperbolic primary mirror, a convex hyperbolic secondary mirror, and a flat imaging device at the focal plane. Given the sensitivity of the plate scale and of the aberrations to the intrinsic parameters and to the relative positions of those three optical elements, special emphasis must be put, firstly on their figure and initial alignment, and secondly on the thermal regulation of the telescope assembly.

The central wavelengths of the five spectral bands ({\it viz.} 215.0, 393.37, 535.7, 607.1, and 782.2\,nm, often referred henceforth as simply `215', `393', `535', `607', and `782') ensue from the following rationale. The photospheric pass bands (`535', `607', and `782') were selected for being quasi free of Fraunhofer lines, so as to account for the pure solar black body and to therefore neglect any overlying structure. They also had to perpetuate valuable historical series, or to initiate new time series if a compelling case could be made. A narrow pass band centered around 535.7\,nm was chosen for its heritage with the Calern measurements although those were carried out in a wider pass band around 540\,nm \cite{Laclare1996SoPh}. A second narrow channel was adopted at 607.1\,nm.
The Precision Solar Photometric Telescope (PSPT) produces images at 607.1\,nm in the red continuum (\url{http://lasp.colorado.edu/pspt_access/}).
It can also be compared with the 590--670\,nm spectral range of the SDS balloon experiment \cite{Sofia1984ApOpt,Egidi2006SoPh}. A third Fraunhofer-line-free passband is centered at 782.2\,nm. It can relate to the `Solar Diameter Monitor' measurements at 800\,nm \cite{Brown1998ApJ} and it offers the steepest limb darkening function (LDF) of the three photospheric channels.

The mechanical design of SODISM, namely its two 5-slot filter wheels, allowed for two additional spectral channels. Pass bands around 215.0\,nm and 393.37\,nm were selected. The first wavelength has a large climatic interest since it is known that the variability of the irradiance in the Herzberg continuum 
(200--242\,nm) 
strongly influences the ozone concentration in the stratosphere \cite{Rozanov2011AGU}. This is why this range is also sampled by the PICARD-PREMOS spectro photometers \cite{Schmutz2009Metro} leading to prospects of cross-calibration with SODISM. High spatial resolution narrow-field observations at 214\,nm have been recorded by the SUFI imager onboard the SUNRISE balloon experiment \cite{Gandorfer2011SoPh}, but 215\,nm full-disc images are not monitored otherwise. The SODISM 215\,nm channel therefore produces data that are very valuable, even on their own.

The other chromospheric channel is centered at 393.37\,nm (Ca\,{\scriptsize II}\,K line, singly ionized calcium, which is magnetically active). This permits to image the low chromosphere \cite{Ermolli2010A&A} and particularly, to observe the plages with an enhanced contrast. The Ca\,{\scriptsize II}\,K line is indeed used by several ground observatories to generate a `plage index' \cite{Bertello2010SoPh}, and more generally, to inform about solar magnetism \cite[and references therein]{Ermolli2009ApJ,Sheeley2011ApJ}. It can be compared with the PSPT experiment.

The above choices about the optical configuration and spectral channels determines the core design of SODISM. Yet, the instrument has been supplemented with a number of important design features that augment its capacities:
\begin{itemize}
\item In closed position, a door covers the front window and protects the whole telescope against solar radiations. Its mechanism allows several opening and closing of the whole telescope, although the operational plan anticipates a single opening that occurred on 21 July 2010, at the beginning of the solar observing program. See section~\ref{SS-Door}.
\item A front window bears on its inner side a reflective and absorbing coating, which divides by $\sim$20 the penetration of radiation inside the instrument. This protection is especially necessary in the ultraviolet and infrared ranges. The purpose is to limit the degradation due to the polymerization of hydrocarbons (UV) and to facilitate the heat management (IR). See section~\ref{SS-FrontWindow}.
\item The whole SODISM assembly is thermally regulated. See section~\ref{SS-TCS}.
\item Two successive filter wheels permit to insert one of the above mentioned spectral filters and other refractive elements (a defocussing lens, and another lens for stellar operations), or to leave the optical path open. See sections~\ref{SS-Filters} and \ref{SS-Wheel}.
\item A $2k\times 2k$ frame-transfer CCD is placed at the focal plane. A shutter mechanism provides it with dark conditions except within the duration of its electronic exposures. The CCD is anti-reflective (AR) coated. See sections~\ref{SS-CameraCCD} and \ref{SS-Shutter}.
\item A radiator and a cold finger remove the heat at the back of the CCD to cool it and decrease its dark signal during normal operations. The low CCD temperature and the dark conditions granted by the shutter provide for the relatively slow readout of the camera ($\sim$22\,s). In parallel, heaters on the cold finger give the possibility to accurately regulate its temperature and to raise it to room temperature for baking out on demand.
\item Three piezo-electric actuators are installed triangularly at the back of the primary mirror (M1). They allow translating it by $\pm$15\,$\mu$m, and tilting it to shift the image of the Sun by $\pm$1\,arcmin. See section~\ref{SS-Piezos}.
\item A ring shape green filter and four prisms are disposed circularly around the external border of the telescope entrance, just behind the front window. This optical subsystem generates four shifted images of the solar disc. Part of those is again deviated by a third folding mirror (M3) that is inclined (by 45$^o$) in order to illuminate four photodiodes at the input of a declutchable image-stabilizing controller which feeds back on the M1 mirror location, {\it via} the three piezos at its back. See section~\ref{SS-StabilizationPointing}.
\item The M3 mirror is hollow in order to let the central main beam and the other parts of the four shifted images reach the selected spectral filter. All five images get focused on the CCD. In particular, the prisms create green secondary images in each corner of the field of view (FOV). Their external shape is determined by the vignetting of the M3 mirror. The wedge angle of the prisms being stable, those corner images can reveal optical evolutions occurring potentially in the rest of the telescope. This is the so-called `internal scale' \cite{Assus2008AN}.
\item Although PICARD's normal attitude aligns the North-South axis of the Sun with the rows of the CCD, the platform is able, on demand, to roll around the optical axis of SODISM in an attempt to separate instrumental effects (such as distortion) from astrophysical quantities (such as solar oblateness). See section~\ref{SS-Ops}.
\item Finally, the platform is also able to quit the nominal solar pointing and overturn in order to point at fields of stars. This capability provides a complementary means to calibrate the optical system, and especially its PSF and its magnification. This is the so-called `external scale'. See section~\ref{SS-Ops}.
\end{itemize}

The details of the SODISM design are now reported in the next sections.

\begin{figure}[h]
	\centerline{\includegraphics[width=.82\textwidth]{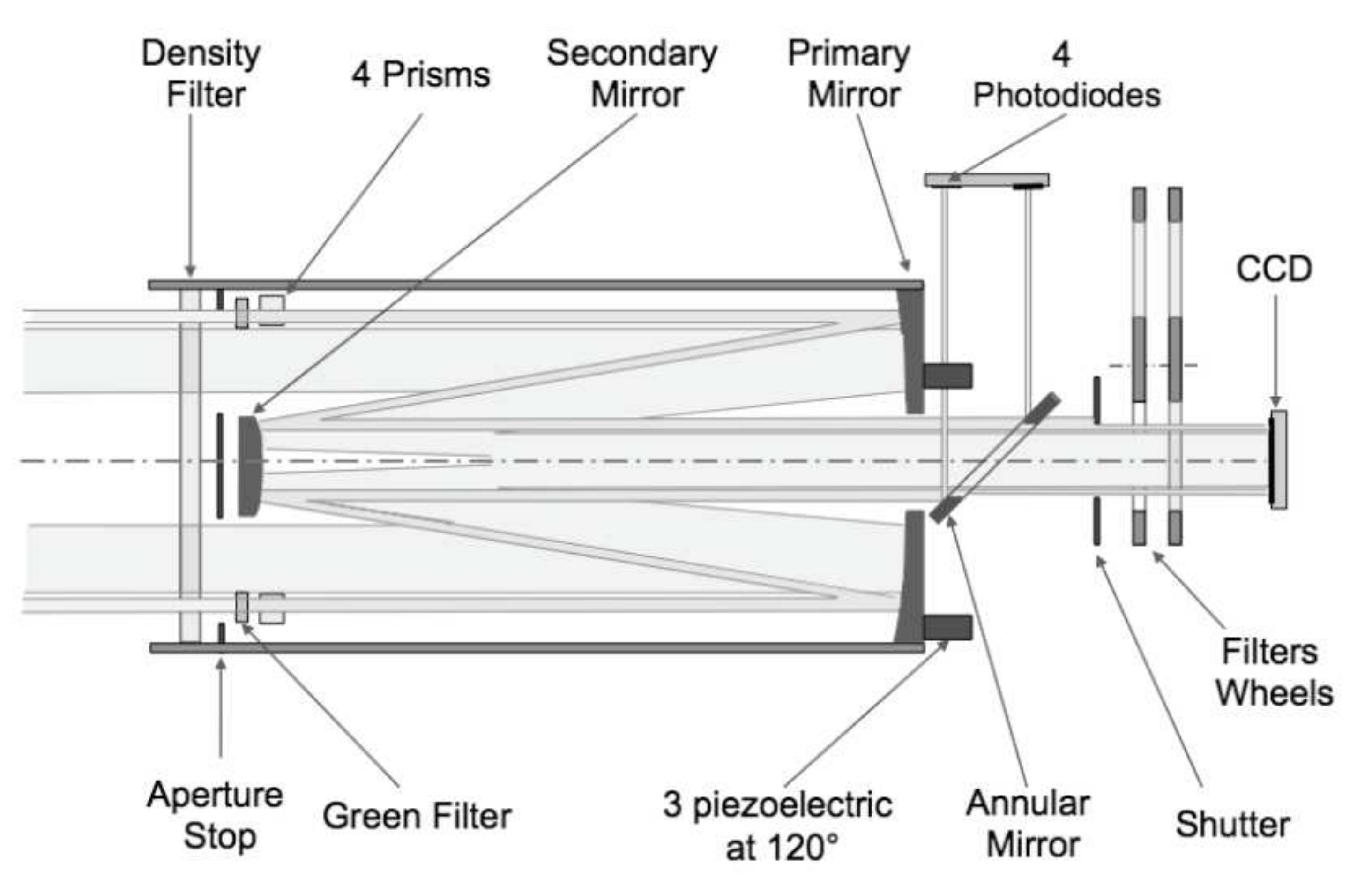}}
	\caption{SODISM main optical path consists essentially of a front window (density filter), a primary mirror (M1), a secondary mirror (M2), interchangeable interference filters, and a CCD. The secondary optical path is represented with a darker shade. It starts with a green filter and four prisms at the level of the pupil. It feeds four corner images as well as an internal pointing and stabilization system. The external portion of the secondary optical path is deviated by the M3 annular mirror onto four photodiodes. Their signal is processed and fed back on three piezoelectrical actuators at the back of the M1.}
	\label{Fig_SODISM_optical_system}
\end{figure}


\section{Optical Design}\label{S-OpticalDesign}

\subsection{Overall configuration}

\begin{figure}[h!]
\centerline{\includegraphics[width=0.70\textwidth]{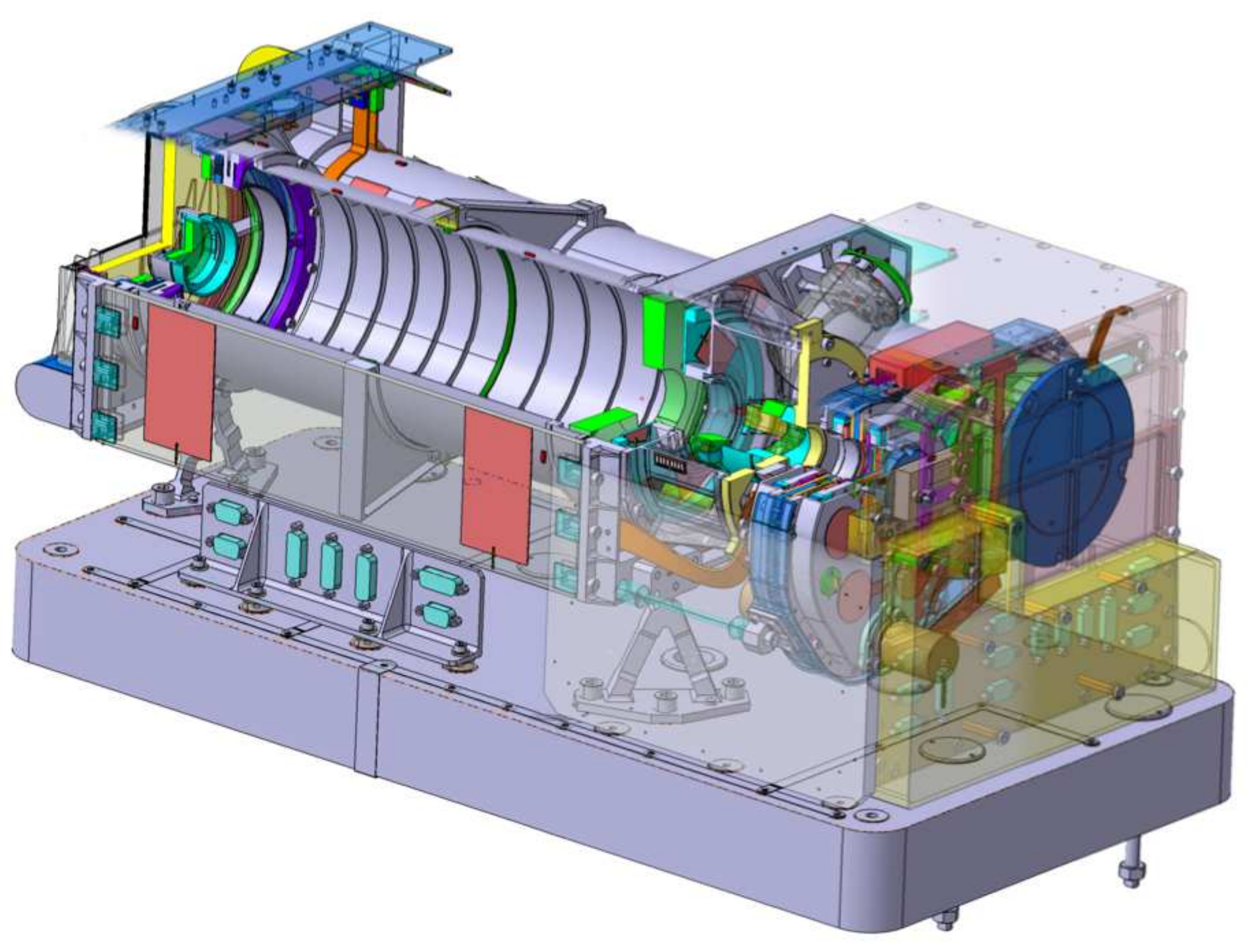}}
\caption{Computer-aided design (CAD) representation of the SODISM telescope. The door, front part, inner baffle, and optical elements can be recognized on the left of the image. The camera, the filter wheels, and the shutter mechanism are represented on the right. A Carbon/Carbon composite monolithic structure (located behind the internal optical baffle) was selected to link the mirror and the CCD because it imparts thermo-mechanical stability thanks to its low coefficient of thermal expansion.
The `Senseur d'Ecartom\'etrie Solaire' (SES) is located inside the monolithic structure.
The SODISM instrument is fixed on a common baseplate. SODISM is held by three feet, which attenuate potential thermo-mechanical stress coming from the interface.}
   \label{Fig_SODISM_telescope}
\end{figure}

SODISM is an 11-cm diameter telescope with a charge coupled device (CCD) at its focal plane. Figure~\ref{Fig_SODISM_optical_system} presents its optical layout, Figure~\ref{Fig_SODISM_telescope} gives a view on its interior, and Table~\ref{T-SODISM characteritics} summarizes the main characteristics of the instrument.

The focal length is 2,626\,mm. The CCD detector array has $2048\times2048$ pixels of 13.5\,$\mu$m pitch. Focal length and CCD format lead to a FOV of $\sim$$36\times36$\,arcmin, and a plate scale of $\sim$$1.06$\,arcsec per pixel. The distance between M1 and M2 is 323.5\,mm, and the distance between M2 and the CCD is 516.8\,mm.

The chosen Ritchey-Chr\'etien configuration minimizes spherical and coma aberrations, but this design entails some field curvature. Since image quality is primarily required at the solar limb, the CCD was consequently to be positioned so as to reach best focus there, {\it i.e.} about 16\,arcmin away from the FOV center.

The pupil is determined by the 89\,mm inner diameter of the ring-shaped green filter. The aperture therefore amounts to f/29.5 and the diffraction limit matches approximately the pixel resolution.
During nominal operations (solar pointed), the Sun is the only significant light source and it almost fills the FOV. This is why no external baffles were foreseen.

As mentioned in section~\ref{S-Choices}, several dioptric components provide spectral selection and flux adjustment. Additionally, together with the M3 mirror and associated subsystems, they enable an intra-SODISM stabilization system and the internal scale. The optical elements will now be reviewed one by one.

\begin{table}[h]
\caption{SODISM main characteristics.}
\label{T-SODISM characteritics}
\begin{tabular}{l l}
\hline
Telescope type & Ritchey-Chr\'etien\\
Focal length 						& 2626\,mm\\
Main entrance pupil 					& 89\,mm\\
Volume 							& 670 (d) x 308 (w) x 300 (h) mm$^{3}$\\
Weight and power consumption		& 27.7\,kg and 43.5\,W\\
Field of view and angular resolution 	& 36\,arcmin and 1.06\,arcsec/pixel\\
EEV-4280 back thinned CCD detector	& $2\times2048\times2048$ and 13.5\,$\mu$m square pixels\\
Data rate 							& 2.2\,Gbits per day\\
\hline
\end{tabular}
\end{table}

\subsection{The front window} \label{SS-FrontWindow}

\begin{figure}[h]
\centerline{\includegraphics[width=0.8\textwidth]{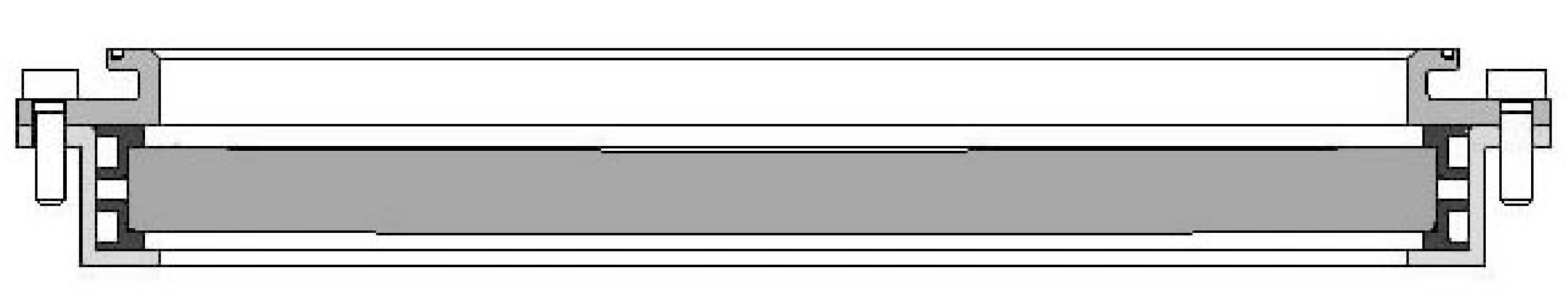}}
\caption{Optomechanical representation of the front window subassembly.}
   \label{Fig_SODISM_density_filter}
\end{figure}

The front window protects the internal SODISM optics from the space environment and limits the solar flux into the instrument, thus restraining heating and degradation.
A schematic view is provided in Figure~\ref{Fig_SODISM_density_filter}.
It has an outer diameter of 126\,mm and a useful diameter of 114.3\,mm. Its central thickness is 8\,mm. It is actually slightly domed, with outwards concavity having a radius of curvature of $\sim$$5\,m$, specified slightly differently for each side and precisely manufactured so as to generate zero optical power. The dome shape is meant to avoid parasitic reflections (`ghosts') which otherwise could have superposed themselves on the CCD at the limb of the main solar disc image.

\begin{figure}[h]
\centerline{\includegraphics[width=1.1\textwidth]{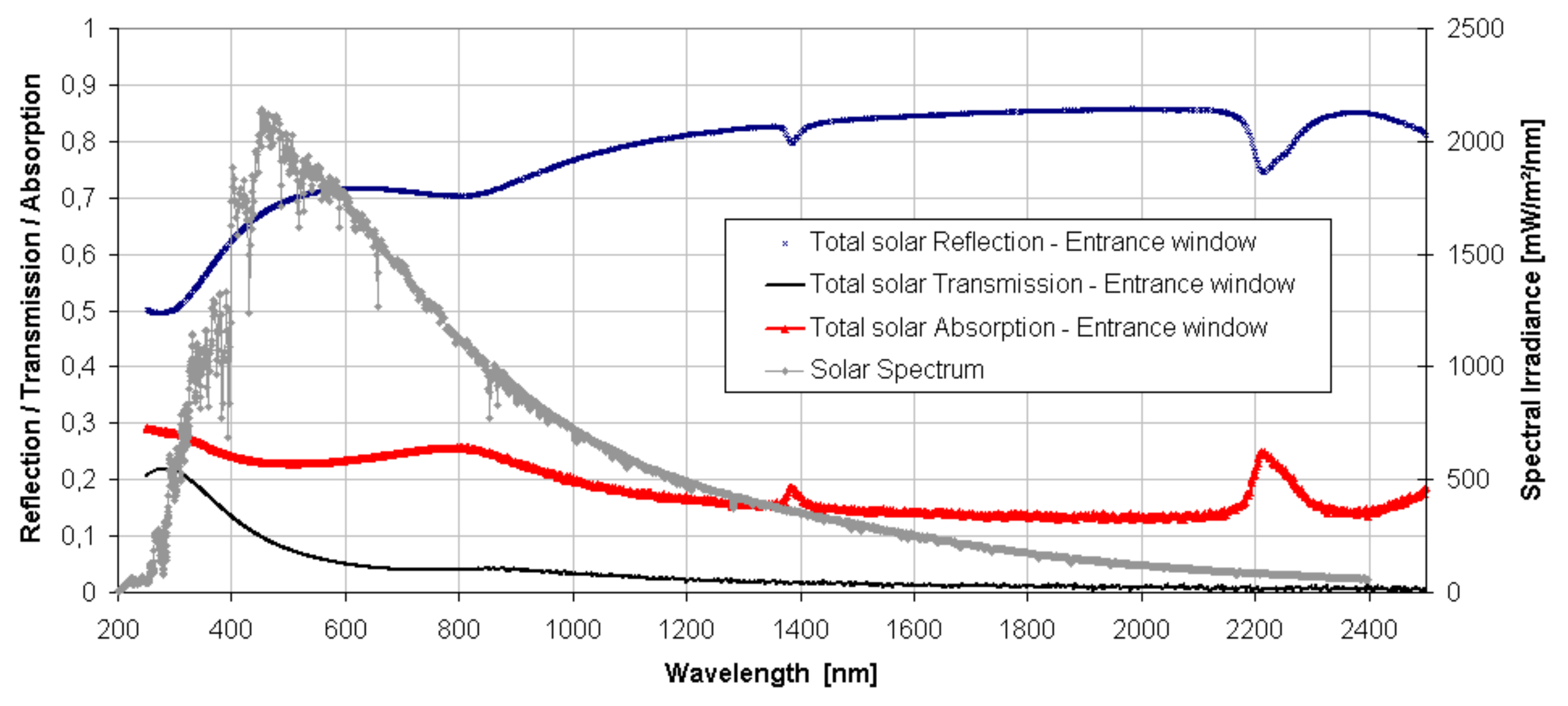}}
\caption{
Reflection, transmission and absorption measurements of the front window. The solar spectral irradiance from 200 nm to 2400 nm represented here in grey has been measured by the SOLSPEC spectrometer.
}
   \label{Fig_SODISM_density_filter_transmission}
\end{figure}

The bulk of the front  window is made of silica (Suprasil). A reflective coating has been deposited on its inner side.
A chromium layer links the  $\sim$$30\,nm$ thick aluminum + $\sim$$30\,nm$ SiO2 layers onto the Suprasil substrate.
It rejects by reflection $\sim$$75\%$ of the solar flux.
Altogether, the front window has a total transmission of 5.8$\%$ and a total absorption smaller than
20$\%$ for the spectral ranges comprised between 250\,nm and 2500\,nm.
Figure~\ref{Fig_SODISM_density_filter_transmission} shows its spectral reflection, transmission, absorption and the solar spectral irradiance measured by the SOLSPEC spectrometer \cite{Thuillier2003}.

Samples of the window have been subjected to thermal cycling (from -40$^{o}$C to +50$^{o}$C), humidity testing (24 hours at +50$^{o}$C with moisture of 90$\%$RH, relative humidity), and to ultraviolet testing (dose equivalent to 6,000 ESH, equivalent Sun hours). No individual test has produced significant effect on the transmission of the samples, but it is difficult to guarantee their cumulated harmlessness.
The front normal emittance has been measured to be around 0.81 and the back emittance to be around 0.08.

The front window is conductively isolated (thermally) from the rest of the instrument. The goal was to exhibit a temperature of at least 20$^{o}$C at the window center. This matches closely the temperature of the front structure and would prevent excessive contamination since the nearby door is at 0$^{o}$C and acts therefore as a cold trap.

\subsection{The mirrors}

All three SODISM mirrors are made of class-2 Zerodur, a stable material with a very low coefficient of thermal expansion (10$^{-7}$ m/m/$^{o}$C). The mirror coatings are multilayer (chrome, aluminum, and SiO2).
Chrome links the substrate with the optical layers.

The optical parameters of the M1 and M2 mirrors are listed in Table~\ref{T-SODISM mirrors}.
The external diameter of the M2 mirror assembly (including a protective mask facing the Sun, on the back of the reflective surface) measures 45.5\,mm.
As a result, the central obstruction of the pupil for the main light track is worth $50\,\%$ in diameter and $26\,\%$ in area.

\begin{table}[!h]
\caption{Characteristics of the flight model (FM) mirrors.}
\label{T-SODISM mirrors}
\begin{tabular}{l l l l l l}
\hline
\#	& Thickness at	& Outside	& Inside	& Radius of		& Conic	\\
	& the vertex 	& diameter	& diameter	& curvature		& parameter	\\
\hline
M1	& 14\,mm		& 121\,mm	& 41\,mm	& -804.61\,mm	& -1.01	\\
M2	& 6\,mm		& 35\,mm	& N/A		& -186.07\,mm	& -1.9382	\\
M3	& 10\,mm		& 46$\times$64.8\,mm	& 29.2$\times$41.2\,mm	& $\infty$		& N/A	\\
\hline
\end{tabular}
\end{table}


\subsection{The spectral filters} \label{SS-Filters}
In addition to the front window, SODISM uses interference filters to adjust the solar flux at the CCD and to determine the narrow pass bands of its spectral channels.
Table~\ref{T-SODISM filters} summarizes the main characteristics of these filters.

\begin{table}[!h]
\caption{Characteristics of the flight model (FM) interference filters.}
\label{T-SODISM filters}
\begin{tabular}{l l l l l}
\hline
Central		& FWHM of	 		& Equivalent 	& Optical	& \\
wavelength 		& the passband		& Transmission at $\lambda$	& thickness					& Provider\\
$\lambda$\,[nm]	& $\Delta\lambda$\,[nm]	& [$\%$]		& [mm]					& \\
\hline
215.00\,$\pm$0.50	& 7.0				& $\sim$3\,10$^{-3}$		& 12.150 		& Acton Research\\
393.37\,$\pm$0.10	& 0.7				& $\sim$8\,10$^{-4}$ 		& 12.274		& Andover\\
535.70\,$\pm$0.05 (Helio) 			& 0.5				& $\sim$4\,10$^{-5}$		& 12.324		& Andover\\
535.70\,$\pm$0.05 	& 0.5				& $\sim$3\,10$^{-4}$		& 12.158		& Andover\\
607.10\,$\pm$0.10	& 0.7				& $\sim$3\,10$^{-4}$		& 12.320		& Andover\\
782.20\,$\pm$0.20 	& 1.6				& $\sim$2\,10$^{-4}$		& 12.296		& Andover\\
\hline
\end{tabular}
\end{table}


  \begin{figure}[h]
   \centerline{\hspace*{0.015\textwidth}
\centerline{\includegraphics[width=.6\textwidth]{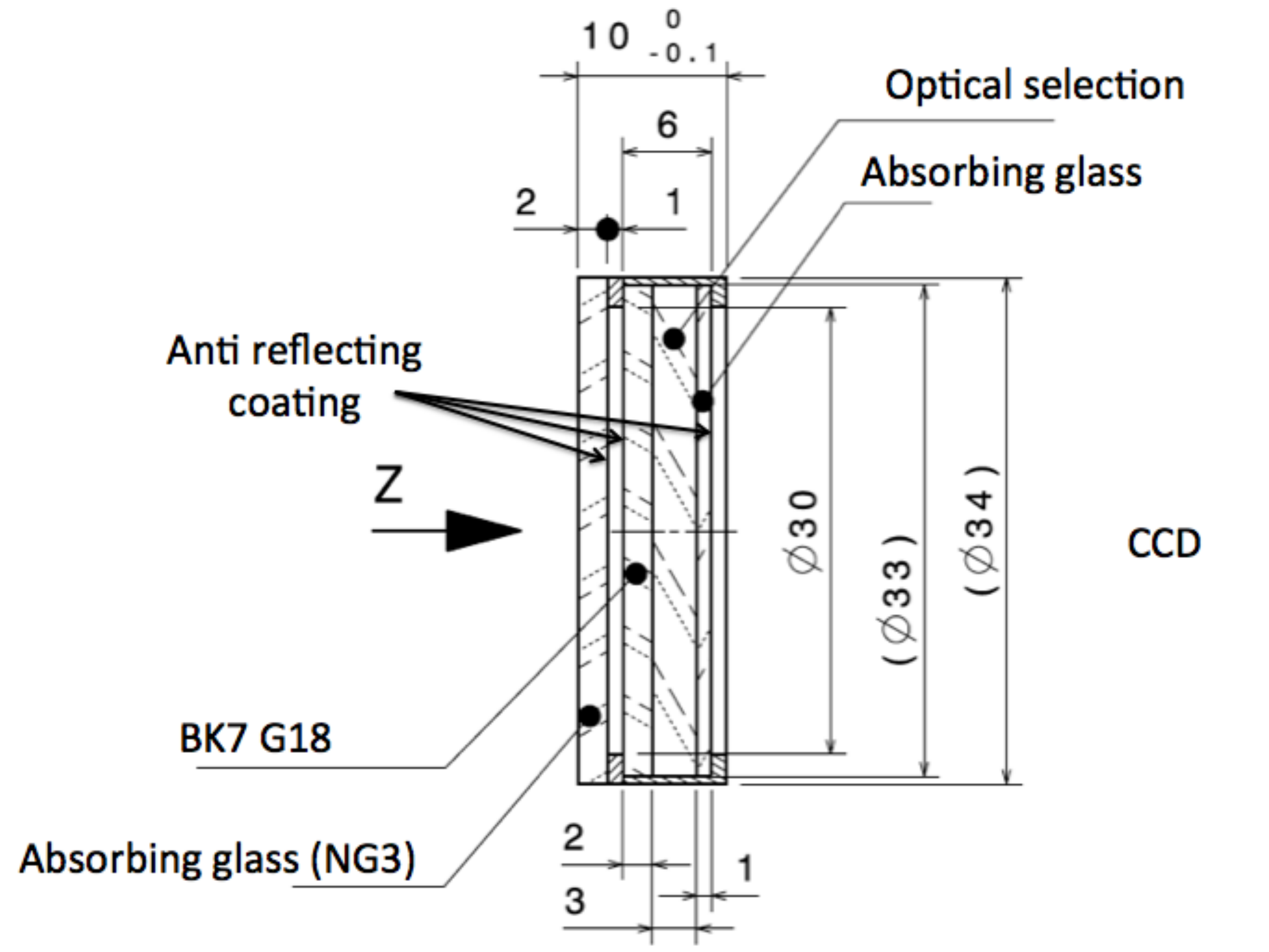}}
              }
     \vspace{-0.35\textwidth}   
     \centerline{\small \bf     
         \hfill}
     \vspace{0.31\textwidth}    
	\caption{Each spectral filter includes an attenuating plate (Inconel front coating and absorbing glass), a filter plate (band pass and blocking coating), a glass plate and an absorbing plate. The three back plates are glued together. Neutral sides bear anti-reflective coatings.}
   \label{Fig_SODISM_filter}
   \end{figure}

For all channels (except at 215.0\,nm), each filter consists of two successive interference filters separated by a small gap. The first one is broadband and the second is narrowband.
These interference filters have been developed by the Andover Corporation (\url{http://www.andovercorp.com}). As to the 215.0\,nm filter, it was procured at Acton Research (\url{http://www.princetoninstruments.com/products/optics/filters/}).

Scientific requirements specify to exclude Fraunhofer lines from the photospheric filter pass bands.
But they are temperature dependent, and typical values are 0.015 nm/$^{o}$C at 500 nm and 0.020 nm/$^{o}$C at 900 nm.
However, the thermal control system (Section~\ref{SS-TCS}) ensures a maximum shift of 0.09 nm over the duration of the mission, which meets the above requirement.

Besides the interference layers, the filters are composed of an Inconel layer deposited on their front surface and of the glasses discussed below.
The Inconel reflects strongly the in-pinging beam in order to tailor the solar flux at the CCD on a channel-per-channel basis.
The reflected beam must normally be sent back to space.

The filters employ the following glasses: BK7G18, GG495, OG590, RG715, NG3, and NG4. 
BK7G18 (from Schott) is used for radiation protection. According to the literature, it displays very small loss of transmission after exposure to 2\,Krad of gamma radiation.
GG495, OG590 and RG715 (from Schott) are colloidally colored long-pass filters.
The CNES space agency demonstrated that, after exposure to 5 Krad of gamma radiation on samples, the peak transmittance was reduced by 0.9$\%$ for GG495, by 0.65$\%$ for OG590, and exhibited no change for RG715.
This test is slightly more severe than the 2\,Krad calculated for a two year mission.
After sample cleaning, the peak transmittance was reduced by 1$\%$ for GG495 and for OG590. There was no effect for the RG715.
Neutral density glasses (NG3 and NG4) exhibit constant transmission over a large spectral range, especially in the visible.

Anti-reflection (AR) coatings have been deposited on the internal and external surfaces of the filter elements. This is to limit the generation of ghost images that would have arisen from reflections between the CCD (despite its AR coating) and the filter elements ahead, or within the filters themselves. It has been verified that resulting ghost images would not exceed a few percent of the main image brightness.

All filter elements were mounted in a black anodized aluminum ring and maintained assembled by epoxies.
In agreement with the ESA-PSS-51 guidelines for spacecraft cleanliness control, all filters exhibit TML (total mass loss) $<$0.1$\%$ and CVCM (collected volatile condensable material) $<$0.01$\%$.

\subsection{The image stabilization and internal pointing system} \label{SS-StabilizationPointing}

\begin{figure}[h]
   \centerline{\hspace*{0.015\textwidth}
               \includegraphics[width=0.40\textwidth,clip=]{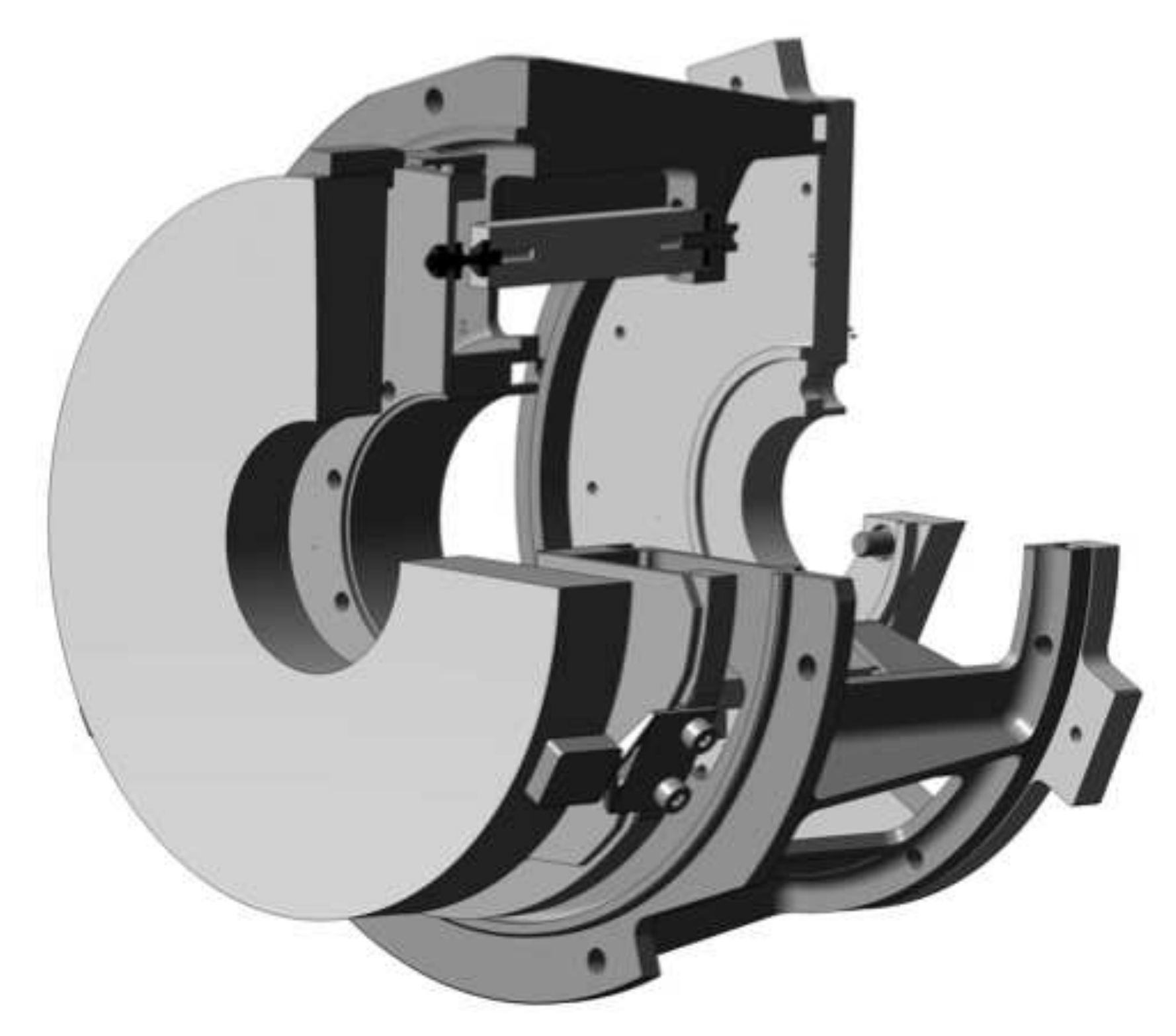}
               \hspace*{0.015\textwidth}
               \includegraphics[width=0.44\textwidth,clip=]{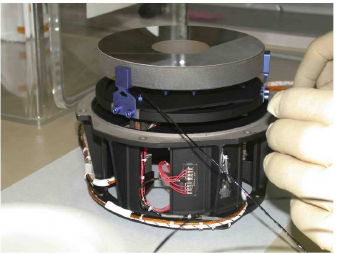}
              }
     \vspace{-0.35\textwidth}   
     \centerline{\small \bf     
      \hspace{0.0 \textwidth}  \color{black}{(a)}
      \hspace{0.39\textwidth}  \color{black}{(b)}
         \hfill}
     \vspace{0.31\textwidth}    
	\caption{The mechanical design of the pointing system is shown in Figure~\ref{Fig_SODISM_mechanism_piezo}-a. Three piezoelectric devices are used to rotate the primary mirror.
A Qualification Model (QM) of the pointing mechanism is shown in Figure~\ref{Fig_SODISM_mechanism_piezo}-b.
	The main design objectives of the mechanism were a pointing accuracy of $\pm$ 0.2 arcsec and a tilt amplitude of $\pm$ 1\,arcmin in both the X and Y directions.
}
   \label{Fig_SODISM_mechanism_piezo}
   \end{figure}

The platform is three-axes stabilized. The AOCS (attitude and orbit control subsystem) is required to provide a pointing accuracy of $\pm$36\,arcsec for the Z-axis of the spacecraft (along the optical path), and a stability of 5\,arcsec/s of the X and Y axis. This level of performance is higher than the performances guaranteed by a standard Myriade platform. Attitude sensing is provided by a star sensor, Sun sensors, a magnetometer, and/or by the telescope; actuation is provided by a set of reaction wheels, and magnetic rods.
The solar images are normally stabilized on the detector by a three-level system:
\begin{enumerate}
\item Two star trackers are used by the platform to achieve pre-pointing within $\pm$0.1\,degree, the platform AOCS feeding back {on the satellite reaction wheels}.
\item A solar tracking system based on a four-quadrant sensor achieves a pointing within $\pm$36\,arcsec, with a maximum drift of 5\,arcsec/s {by feeding back also on the PICARD reaction wheels}. This system is declutchable on demand, typically during deep eclipses.
\item A fraction of each four prism-generated image between 500\,nm and 600\,nm is sent by the M3 folding mirror to a four-diode system. Their signals are processed onboard and the output of the associated control loop allows feedback on the three piezo actuators at the back of the M1 mirror, thereby positioning and stabilizing the solar image on the CCD with a precision of $\pm$0.2\,arcsec. It is represented in Figure~\ref{Fig_SODISM_mechanism_piezo}. This system is also declutchable on demand, typically during weak eclipses or to assess its performance.
\end{enumerate}


\subsubsection{The piezo-electric actuators} \label{SS-Piezos}
The piezoelectric actuator technology has been used in various instruments, notably on SOHO and ROSETTA. For SODISM, piezoelectric actuators, procured from ``Cedrat Technologies'' with reference PPA40M, were modified to reach a higher mechanical preload and to include piezoelectric ceramics. Preloading such components is essential, as they cannot bear any tensile stress. The parallel pre-stressed actuator is a preloaded stack of low voltage piezoelectric ceramics.
Strain gauges are used for each piezoelectric device for repeatability of the positioning.

Each component has its own characteristics, and, for 170 V applied, the FM devices `1', `2', \& `3' demonstrated peak-to-peak displacement amplitudes of 50.04, 50.39, and 50.84\,$\mu$m respectively.
These three piezo devices have been arranged triangularly at the back of the M1 mirror.
They allow to translate it along the Z optical axis of the telescope, thus enabling a `piston mode', and to tilt it perpendicularly, around the X and Y axis, in view of the functions that are explained below in section~\ref{SS-InternalPointingInterest}.

\subsubsection{The four photodiodes and the feedback loop} \label{SS-Stabilization}
The mentioned four-diode system is located aside the main beam track, thanks to the M3 folding mirror. The photodiodes reference is API SD9973.
By design, any signal unbalance between pairs of detectors indicates a departure away from a situation whereby the solar disc would be near the center of the FOV. More generally, the needed unbalance can be computed and commanded {\it via} the piezo actuators in order to translate the solar image by a desired amount in the X and Y directions. Further, during an exposure, the unbalance configuration is controlled by a feedback loop in order to stabilize the solar image with respect to the CCD, thereby avoiding kinematic blur.
The precision is expected to be better than $\pm$0.2\,arcsec rms for the duration of an exposure.

This subsystem therefore provides SODISM with an internal pointing as well as with an image stabilization capability.
Solar images are recorded every minute. The pointing mechanism is therefore permanently operational, but not continuously operated. Outside exposure times, the piezo actuators are supplied with 60\,Volts, which correspond to their mean position. The strain gauges are driven uninterruptedly to reach a stable thermal behavior. Image stabilization is activated by the feedback loop two-seconds before shutter opening, and of course, during the whole exposure.
The piezos thereafter return to their 60\,V position. 
This subsystem delivers estimates of the angular displacement of the primary mirror (so-called P5X and P5Y), and of its translation (so-called P4). These values are embedded in the housekeeping (HK) data.

\subsubsection{The benefits of internal off-pointing}\label{SS-InternalPointingInterest}

Beyond its original motivation, {\it viz.} solar disc centering and image stabilization, the described subsystem bestows SODISM with the possibility to tilt its M1 mirror around the X and Y axis, and to translate it along the Z axis. This offers at least three applications that will be briefly described below. Further report will appear in other papers.

The first benefit derives from the creation of a dataset where solar images are translated to different locations in the FOV. Several such datasets have been programmed and recorded. For some, the Sun center was made to follow a cross-shaped trajectory; for others, the Sun centers were positioned on an irregular grid. Using a recognized method \cite{Kuhn1991PASP}, it has  then been possible to estimate the flatfield of the SODISM instrument.
As the mechanical range is rather small, mainly the small scale component is recovered.
By repeating the procedure after some months, the evolution of the flatfield can be monitored in time. It could for example be shown that the `782' flatfield is strongly dominated by an interference pattern in the surface layer of the CCD. More generally, different imperfections to the flatfield are extracted and this will be discussed elsewhere.

A second benefit arises from the ability of the above datasets to also enable the reconstruction of an unwanted ghost image that forms on top of the solar disc image \cite{etcheto2011SPIE}. Indeed, varying the M1 tilt so that the ghost model remains in agreement with the ghost observation provides a way to estimate the unknown optical parameters that lead to the formation of a ghost image. This study will be reported in another paper.

Thirdly, in-phase driving of the piezos enables a `piston mode', which varies the M1--M2 distance and hence modulates the level of focus of the telescope. This enables a phase-diversity approach, which was however not foreseen beforehand. The motion amplitude is in effect limited, and the desired analysis is not yet substantiated. Due to this amplitude limitation, the pointing and stabilization system presented above is unable to correct for any significant departure from perfect focus.

\subsection{Point Spread Function}

\begin{figure}[h]
	\centerline{\includegraphics[width=1.2\textwidth]{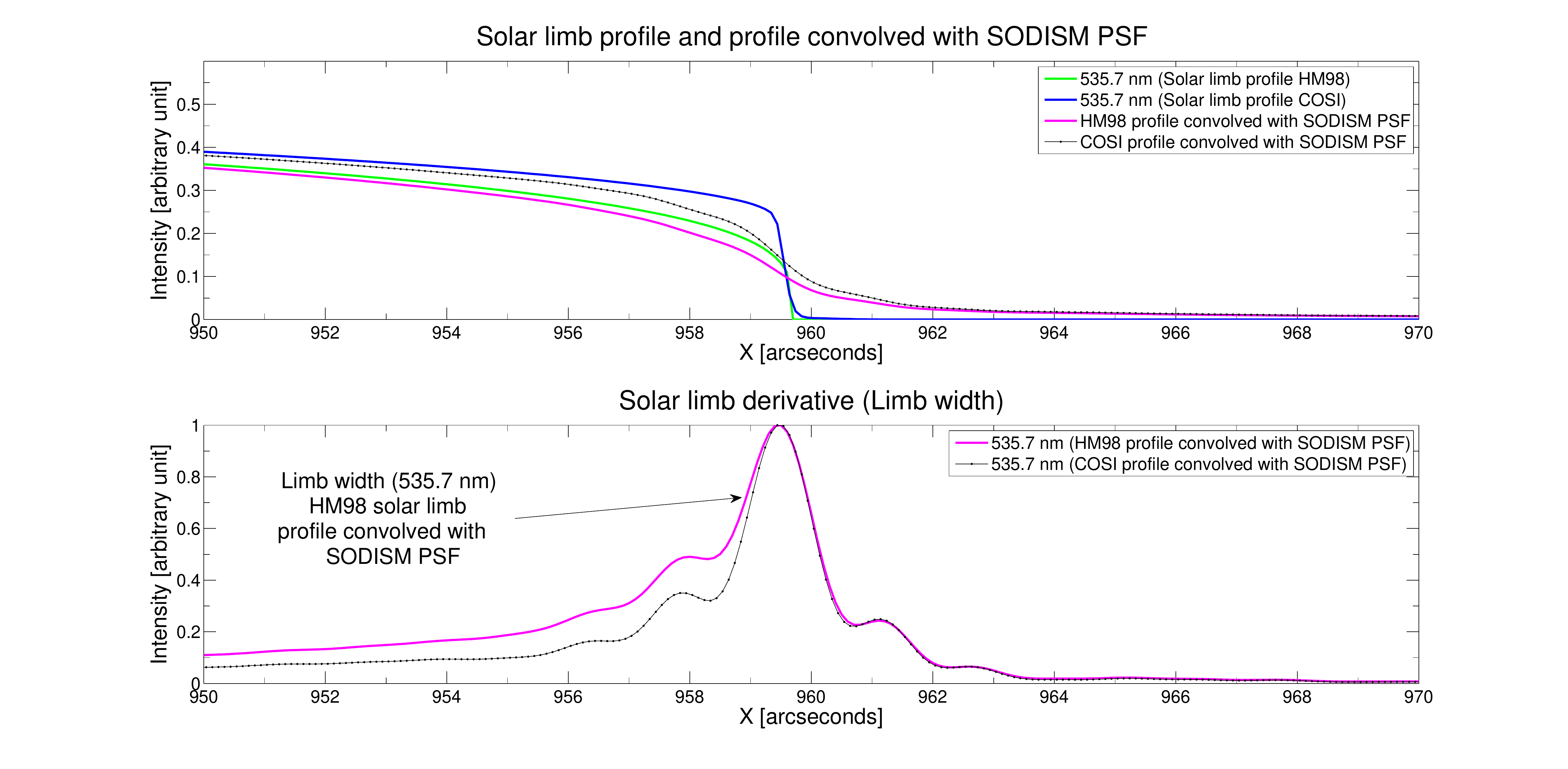}}
	\caption{The top panel shows the limb darkening function (LDF) at 535.7\,nm for the HM98 (empirical power law model extrapolated to the limb) and COSI models, in green and blue respectively. Their convolution with the nominal PSF of SODISM (dominated by diffraction) is over-plotted in pink and black. This PSF has been estimated by an optical model of SODISM under the Code V software. The bottom graph displays the first derivative of these observables, using the same color code. It can be verified visually that the nominal limb width is less than two arcsec and that the inflection point appears to be unequivocally defined for either radial limb profile.}
	\label{Fig_SODISM_limb_psf}
\end{figure}

The SODISM Point Spread Function (PSF) and its effect on the solar limb have been studied for the nominal optical configuration, wherein the instrument is diffraction-limited.
Indeed, SODISM design and dimensions are such that the size of the Airy disc of diffraction is about one arcsec in the bluest case (215.0\,nm), which is still larger than any of the expected aberrations.
The limb darkening function (LDF) of HM98 \cite{Hestroffer1998AA} and COSI \cite{Haberreiter2008AA,Shapiro2010AA} have been convolved by the theoretical PSF of the telescope and the results at 535.7\,nm are represented in Figure~\ref{Fig_SODISM_limb_psf}.
The first derivative of the limb is spread over about two arcsec, and the location of the inflection point (maximum of the first derivative) appears to be well defined for both LDF models.

The study of the PSF in non nominal configurations (slight modifications in the figure or in the alignments of the mirrors, temperature gradients in the front window) is ongoing. Some outcomes are provided in the next section.

\subsection{Optical tolerances}

The accuracy requirements in the positions and figures of the various optical elements are high, for the effect of aberrations to remain small compared to the size of the Airy disc.

\begin{table}[!h]
\caption{SODISM optical elements positioning error budget. The optical path is along the Z direction. T are translations, and R are rotations.}
\label{T-SODISM optical position}
\begin{tabular}{l l l l}
\hline
Optical element		&Tx						&Ty						&Tz					\\
\hline
Front window			&+0.245/-0.155 mm			&+0.245/-0.155 mm			&+0.015/-0.3 mm		\\
Secondary mirror M2	&+0.106/+0.007 mm			&+0.106/+0.007 mm			&$\pm$0.219 mm		\\
Primary mirror M1		&+0.111/+0.031 mm			&+0.111/+0.031 mm			&$\pm$0.24 mm		\\
Filter					&+0.55/+0.133 mm			&+0.55/+0.133 mm			&+0.152/-0.192 mm		\\	
CCD					&$\pm$0.5 mm				&$\pm$0.5 mm				&$\pm$0.1 mm			\\
\hline
\hline
Optical element		&Rx						&Ry						&Rz					\\
\hline
Front window			&$\pm$2 arcmin			&$\pm$2 arcmin			&--\\
Secondary mirror M2	&$\pm$7 arcmin			&$\pm$7 arcmin			&--\\
Primary mirror M1		&$\pm$3 arcmin			&$\pm$3 arcmin			&--\\
Filter					&$\pm$16 arcmin			&$\pm$16 arcmin			&$\pm$20 arcmin\\	
CCD					&$\pm$12 arcmin			&$\pm$12 arcmin			&$\pm$5 arcmin\\
\hline
\end{tabular}
\end{table}

\subsubsection{Positioning errors}

After manufacturing the mechanical parts, we have established an error budget in terms of the translations and rotations of the main optical elements of SODISM.
Table~\ref{T-SODISM optical position} summarizes this error budget.

The requirement for the filters is a Wave Front Error (WFE) smaller than $\lambda$/4 and a prism that must be less than 30 arcsec. A tilt of 16 arcmin between the filters and the optical axis is tolerated on the basis of its spectral effect. It is observed later that it generates an unwanted parasitic reflection that is discussed elsewhere \cite{etcheto2011SPIE}. The tolerance on detector decenter with respect to the telescope axis is about 1 arcmin ($\sim$0.5\,mm).

\subsubsection{Stability budget for the diameter measurements of the Sun}

Solar diameter measurements are impacted if any of the optical element translates or rotates more than acceptable.
A modeling based on a solar diameter defined by the {\it locii} of the inflection point at the solar limb, and based on optical modeling estimates the effect of the various motions on the measurement. This can then be converted into an error budget for the mechanical positioning, which is provided in Table~\ref{T-SODISM radius simu}. The consequences on the solar radius being of the order of few milli arcsec during one orbit, SODISM appears robust against changes of the order of few microns or few degrees of temperature.


\begin{table}[!h]
\caption{Simulated effects of the telescope prescription on the estimated Solar radius.}
\label{T-SODISM radius simu}
\begin{tabular}{l l l l}
\hline
Error Source				& Nature of the error 		& Effect on the radius		\\
\hline
T$^o$ change in the telescope	& Distance M2-M1 along Z		& +2.2 mas/+1$\mu$m		\\
T$^o$ change in the telescope	& Distance CCD-M1 along Z	& +1.6 mas/+1$\mu$m		\\
T$^o$ change in the CCD		& CCD geometry			& -1.88 mas/+1$^{o}$C		\\
T$^o$ change in the M1 mirror	& Curvature of M1 			& +0.08 mas/+1$^{o}$C		\\
T$^o$ change in the M2 mirror	& Curvature of M2			& +0.09 mas/+1$^{o}$C		\\
T$^o$ gradient in the M1 mirror	& Curvature of M1		 	& +0.96 mas/+1$\mu$m		\\
T$^o$ gradient in the M2 mirror	& Curvature of M2			& +4.54 mas/+1$\mu$m		\\
T$^o$ gradient in the front window	& Axial gradient			& +0.15 mas/+1$^{o}$C		\\
\hline
\end{tabular}
\end{table}

\subsubsection{Effects of the misalignments on the limb width}

Another way to look at the effect of misalignments is presented in Table~\ref{table_limb_width_Effect}. In this table, the increase of the thickness of the limb is shown to depend on the chosen type of radial limb profile HM98 \cite{Hestroffer1998AA}. Moreover, small displacements of the M2 mirror along the Z axis appear to modify significantly the radial shape of the limb. This makes the M2 positioning critical.


\begin{table}[!h]
\caption{Limb width effect (535.7 nm).}
\label{table_limb_width_Effect}
\begin{tabular}{l l l}
\hline
Displacement type		& Displacement		 				&Limb width effect \\
					& amount								& (HM 98) [arcsec] \\
\hline								
CCD defocalisation		& 2\,mm								& 1.54 \\
Tz					& 3\,mm								& 4.98 \\
					& 4\,mm								& 8.52 \\
					& 5\,mm								& 10.87 \\
\hline									
M2 mirror displacement	& 10\,$\mu$m							& 0.1 \\
Tz					& 20\,$\mu$m							& 0.44 \\
					& 50\,$\mu$m							& 1.27 \\
					& 100\,$\mu$m							& 8.66 \\
\hline				
M2 mirror shift			& 100\,$\mu$m							& 0.11 \\
Tx or Ty				& 200\,$\mu$m							& 0.2 \\
					& 500\,$\mu$m			 				& 0.33 \\
					& 1000\,$\mu$m						& 0.76 \\
\hline
\end{tabular}
\end{table}

\subsubsection{Effects of gradients in the front window}

We  analyze the effect of a radial thermal gradient in the front window.
A radial thermal gradient has two influences: bending symmetrically the radii of curvature of both faces inducing optical power in the window, and creating a refractive index gradient, which also induces optical power.
The A.~Gullstrand formula gives the focal length of a diopter having a temperature gradient:
	\begin{equation}
	\label{eq:alpha}
		f^{\prime}_{F}=-\frac{1}{2}	          \frac{n_{F}}{n_{F}-1}			\frac{R_{F}^{2}}{e_{F}(\beta_{F}+n_{F}\alpha_{F}){\Delta}T_{F}}\,		
	\end{equation}
where $f^{\prime}_{F}$ is the focal length of the window exhibiting a temperature gradient, $n_{F}$ is the refractive index (for Suprasil it is 1.46008 at 546.07\,nm), $R_{F}$ is the semi diameter of the window (63\,mm), $e_{F}$ is its thickness (8\,mm), ${\beta}_{F}$ is its coefficient of refractive index thermal variation (10.2 ppm/$^{o}$C at 546.07 nm), ${\alpha}_{F}$ is its coefficient of thermal expansion (5.5 10$^{-7}$ m/m/$^{o}$C), and ${\Delta}$$T_{F}$ is the gradient across the window from its edge to the center.

The impact of various center-to-edge temperature gradients in the front window are shown in Table~\ref{T-SODISM radius fen simu}. It can be concluded that such gradients have non negligible consequences on the measurement, contrarily to gradients in the mirrors (the analysis of which being not reported here). Gradients vary during an orbit (variations of outgoing longwave radiations).

\begin{table}[!h]
\caption{Simulated effect of different center-to-edge temperature gradients on the solar radius (simple theoretical approach).}
\label{T-SODISM radius fen simu}
\begin{tabular}{l l l l}
\hline
Radial gradient & ${\Delta}$$T_{F}$=+1$^{o}$C & ${\Delta}$$T_{F}$=+5$^{o}$C & ${\Delta}$$T_{F}$=+10$^{o}$C \\
\hline
  Focal length $f^{\prime}_{F}$ & -75 km & -15 km & -5 km\\
  Geometrical effect on the radius & -3.6 mas & -11.5 mas & -26 mas\\
\hline
\end{tabular}
\end{table}

\section{The CCD and its Electronics} \label{SS-CameraCCD}

\subsection{The Charge Coupled Device (CCD)}

SODISM uses a 42-80\,CCD from E2V. This imaging sensor is a frame-transfer matrix of $2\times2048\times2048$ pixels, 13.5 $\mu$m square pixel size. The CCD area is split into two zones: one $2048\times2048$ image zone (IZ) where the signal is accumulated during integration, and a $2048\times2048$ memory zone (MZ) where the signal is stored during readout. The frame-transfer design would allow shutterless operations and exposing while reading out. However neither of these two features is used and the main rationale for choosing this device has been synergy with the contemporaneous COROT astrometric program \cite{Lapeyrere2006MNRAS}. The main characteristics of the SODISM CCD are given in Table~\ref{T-SODISM filters CCD}.

The Flight Model (FM) CCD is thinned and back-illuminated. This technology improves the sensitivity, especially in the MUV/NUV ranges. The sensor has a high quantum efficiency (QE) in the [370--950]\,nm spectral range.

It operates under MPP (multi pinned phase) mode. Associated with a temperature regulated at about -10$^{o}$C, the MPP technology reduces the dark signal to an acceptable level.

\begin{table}[!h]
\caption{Some characteristics of the SODISM CCD and associated camera.}
\label{T-SODISM filters CCD}
\begin{tabular}{l l l l}
\hline
Photo Response Global Non Uniformity 		& 	At 400 nm		&	12.10$\%$ \\
									&	At 550 nm		&	10.10$\%$ \\
									&	At 650 nm		&	5.40$\%$	 \\
									&	At 750 nm		&	9.00$\%$	 \\
									&	At 900 nm		&	11.50$\%$ \\
\hline
Quantum efficiency						&	At 300 nm		&	14.60$\%$ \\
									&	At 350 nm		&	18.90$\%$ \\
									&	At 400 nm		&	53.90$\%$ \\
									&	At 500 nm		&	81.20$\%$ \\
									&	At 650 nm		&	86.90$\%$ \\
									&	At 800 nm		&	63.50$\%$ \\
\hline
Charge Transfer Efficiency (CTE)			&	Vertical		&	0.999999	 \\
Maximum local dark signal				&	-40$^{o}$C	&	1.1 $e^{-} pixel^{-1} s^{-1}$  \\
Maximum local dark signal				&	-7.2$^{o}$C	&	$\sim$ 4.0 $e^{-} pixel^{-1} s^{-1}$ \\
Maximum local dark signal (calculated)		&	+20$^{o}$C	&	6493 $e^{-} pixel^{-1} s^{-1}$ \\
\hline
\end{tabular}
\end{table}

\subsection{Camera Electronics}

The camera electronics has two functions. First, it supplies the CCD and the shutter mechanism with various bias voltages and it sequences them with the needed clocks.
The reset drain voltage (V$_{RD}$) was set at 8.4\,V.
The substrate voltage (V$_{SS}$) was set at 0.0\,V.
The output gate voltages (V$_{OG1}$ \& V$_{OG2}$) were set at -6.6\,V and -5.6\,V, respectively.
The output drain voltage (V$_{OD}$) was set at 21.7\,V.

Secondly, it amplifies and converts the low analog signals that is read out from the two output ports of the CCD into a stream of digital numbers that is directed to the computing unit Picard Gestion Charge Utile (PGCU, see section~\ref{SS-Ops}), where the image is formed, processed, and most often, compressed. The camera was designed by the `Service d'A\'eronomie' of the CNRS.

It should be noted that the shutter exposure occurs strictly {\it within} the duration of the electronic integration, which always exceeds the former by 0.4\,s for this reason. Therefore, the dark signal integration time is longer than the illuminated exposure time by this amount. The shutter exposure duration can be programmed to take any value between 0.5\,s and 16\,s by steps of 0.1\,s. The integration time goes in parallel from 0.9\,s to 16.4\,s. When the integration is over, given the frame-transfer structure of the CCD, the signal that has accumulated in the image zone is rapidly transferred to its memory zone. This takes only 20\,ms and occurs anyway under dark conditions.

The second main function of the camera deals with signal amplification and analog-to-digital conversion (ADC).
The gain of the preamplifier is $\sim$0.0045 V/mV.
In order to avoid negative values that the ADC would not be able to convert, an offset is added to the signal before it is fed to the ADC.
Its value is 32\,mV, which is in the [840--850]\,DN range for the `left' port of the CCD, and in the [810--820]\,DN range for its `right' port. These values are systematically estimated onboard by calculating  the average and the standard deviation of an under-scan area, and transmitting the (rounded) result of these as part of the HK (housekeeping).
The ADC then operates on 16\,bit, but for the lossy compressed images only 15\,bit are retained.
The pixel conversion cadence is 100\,kHz, hence a CCD line rate of 10.2\,ms per line, and a total readout duration of 22\,s for a full frame (on the two CCD channels simultaneously).
The overall gain of the readout chain has been estimated to be $\sim$1.2\,ADU/e$^{-}$, {\it i.e.} K $\cong$ 0.83\,e$^{-}$/ADU (when the image is coded on 16 bits), with a discrepancy of about 2$\%$ between the two CCD ports.
The readout noise (RN) is worth $\sim$15\,e$^{-}$\,RMS.

The nominal SODISM cadence is one image per minute. This is also the maximal image cadence. It includes sequentially:
\begin{enumerate}
\item the filter wheels positioning (except for dark signal recording),
\item the exposure, and associated shutter operations in the solar imaging case,
\item the CCD readout,
\item the different onboard processing.
\end{enumerate}


\section{Mechanisms and Mechanical Design}

\subsection{The front door} \label{SS-Door}
The door at the entrance serves to protect the front window against contamination during the outgassing phase, between 15 June 2010 (launch date) and 21 July 2010 when the door has been fully opened.
It weights overall 0.98\,kg and a large plate (168$\times$250$\times$54\,mm) covers or clears simultaneously the front window and the nearby SES (`Senseur d'Ecartom\'etrie Solaire', {\it viz.} the 4-quadrant sensor allowing fine solar pointing by the PICARD platform).
Its mechanism consists of a stepper motor from THALES AEM, a gear-box designed by CNRS and THALES AEM, two pairs of preloaded ball bearings from ADR (reference WX9ZZT4DOK3682), two ABB micro-switch (reference 6932) and a pin puller P25 provided by TiNi Aerospace.
It keeps the door locked closed during the launch phase. In space, it is unlocked as soon as power is available.
The gear-box is 5-stage epicycloidal  and has a ratio 1166:1. It has been specially designed for SODISM so as to maintain a permanent torque on the door even when the motor is not powered.

The door can be commanded into three positions: `closed', `partially closed', and `opened'.
The door is anticipated to always remain in open position during science operations. Once in space, its mechanism was operated to contribute to the outgassing phase of the instrument. Six days after launch, the door was first set into the `partially closed' position for about a month. Then, on 21 July 2010, it was set into the `opened' position.
Late in the mission, it must be again operated.
The door has been qualified to survive a minimum of 1,225 cycles (stepper gear-motor). 
\subsection{The filter wheel mechanism} \label{SS-Wheel}

Two filter wheels are located behind the shutter, the second one being $\sim$15\,mm ahead of the CCD.
Tele-commands can program angular positions for each of them. The main configurations are summarized in Table~\ref{T-SODISM filters position}.
Their main function is to select a channel, {\it i.e.} a spectral domain of observation for SODISM.
The outer diameter of the various filter elements is 34\,mm, but their area is actually limited by the diaphragming effect of their mount (32\,mm).
Each wheel accommodates five elements.
On the second wheel, two of them are {\it not} narrow band filters, nor an open aperture.
The first one is a defocusing lens that can be inserted in cascade with the filters of the first wheel ahead, in order to {\it e.g.} contribute to assessing the flatfield of the CCD. The second one is a diopter that has the same optical depth as the spectral filters, and which thus needs to be used (alone) during the stellar pointing when the low-transmitting filters have to be removed.

\begin{figure}[h]
   \centerline{\hspace*{0.015\textwidth}
\centerline{\includegraphics[width=0.55\textwidth]{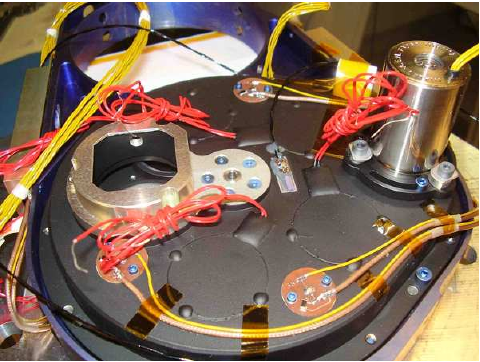}}
              }
     \vspace{-0.35\textwidth}   
     \centerline{\small \bf     
         \hfill}
     \vspace{0.31\textwidth}    
	\caption{One of the two filter wheels during the integration of the telescope.}
   \label{Fig_SODISM_mechanism}
   \end{figure}

The wheels are actuated by stepper motors.
The positioning time is less than 5.6\,s. This is short, as compared to the nominal 1\,min cadence of the camera.
The precision of the positioning of a filter around the optical Z axis is $\pm$20\,arcmin.
To minimize the risk of contamination, Molybdenum disulfide (MoS2) is used as the (solid) lubricant on the ball bearings of the wheels.
Those mechanisms (Figure~\ref{Fig_SODISM_mechanism}) have had to operate in vacuum at 293 Kelvin during a 6-month ground-test, and then for 2 years at least in space.
The filter wheels have been qualified to survive a minimum of 444,400\,cycles (one cycle is a two-way motion).


\begin{table}[!h]
\caption{Usual configurations of the filter wheels.}
\label{T-SODISM filters position}
\begin{tabular}{l l l l}
\hline
\rule[-1ex]{0pt}{3.5ex}  Configuration purpose & Filter wheel \#1 & Filter wheel \#2\\
\hline
\rule[-1ex]{0pt}{3.5ex}	215\,nm observations				&	Filter `215' (1-5)	&	Open (2-2)\\
\rule[-1ex]{0pt}{3.5ex}	393.37\,nm observations				&	Open (1-1)		&	Filter `393' (2-4)\\
\rule[-1ex]{0pt}{3.5ex}	535.7\,nm helioseismic observations	& 	Open (1-1)		&	Filter `535H' (2-1)\\
\rule[-1ex]{0pt}{3.5ex}	535.7\,nm observations				&	Filter `535D' (1-2)	&	Open (2-2)\\
\rule[-1ex]{0pt}{3.5ex}	607.1\,nm observations				&	Filter `607' (1-3)	&	Open (2-2)\\
\rule[-1ex]{0pt}{3.5ex}	782.2\,nm observations				&	Filter `782' (1-4)	&	Open (2-2)\\
\rule[-1ex]{0pt}{3.5ex}	Blurred 215\,nm observations			&	Filter `215' (1-5)	&	Defocusing lens (2-5)\\
\rule[-1ex]{0pt}{3.5ex}	Blurred 535.7\,nm observations		&	Filter `535D' (1-2)	&	Defocusing lens (2-5)\\
\rule[-1ex]{0pt}{3.5ex}	Blurred 607.1\,nm observations		&	Filter `607' (1-3)	&	Defocusing lens (2-5)\\
\rule[-1ex]{0pt}{3.5ex}	Blurred 782.2\,nm observations		&	Filter `782' (1-4)	&	Defocusing lens (2-5)\\
\rule[-1ex]{0pt}{3.5ex}	Stellar observations					&	Open (1-1)		&	Diopter (2-3)\\
\rule[-1ex]{0pt}{3.5ex}	Dark signal observations				&	Any				&	Any\\
\hline
\end{tabular}
\end{table}

\subsection{The shutter} \label{SS-Shutter}

SODISM mechanical shutter is an electro-programmable device with a large 35\,mm aperture.
VS35 types were procured from Vincent Associates / Uniblitz (\url{http://www.uniblitz.com}). This reference has been used in other space-borne applications in the past.
Its space heritage includes the SOLSE experiment in 1997 and SOLSE-2 in 2003 aboard the Space Shuttle, 
and the 25\,mm version flew to Comet Halley aboard the Vega Probes in 1986. 
The uncased version allowed flexibility for integrating the VS35 shutter in SODISM.

Commercial Off-The-Shelf (COTS) mechanisms may not fulfill demanding space applications because they do not survive the launch load or might stop functioning in the hostile space environment.
A dedicated program thus adapted the COTS design in order to increase its robustness. 
The FM shutter has thereafter been qualified to survive a minimum of 1,329,560 opening and closing operations.

The electronics associated to the shutter of SODISM permits exposure times comprised between 0.5 second and 16 seconds.
The total time it needs to switch from closed to opened or back is less than 30 milliseconds.
The exposure time is measured with an accuracy of $\pm$36 microseconds 
in agreement with the needs of the science objectives, in particular those of the helioseismology.
This information is telemetered as part of the housekeeping (HK) information.

\subsection{The mechanical model} \label{SS-MechanicalModel}

The telescope requires a very high level of stability to reach its science objectives, and especially to obtain the needed accuracy and precision on solar diameter measurements.
For this reason, a number of mechanical options were taken and validated by a structural (Finite Element) model and by a thermo-mechanical model (Section~\ref{SS-ThermalModel}).

The SODISM telescope uses the Carbon/Carbon \cite{Meftah2011SPIE}, Invar, and Titanium materials for their respective mechanical value.
Following a sine test, the telescope first eigen mode is 182 Hz, in agreement with the structural model which gives 163 Hz.

The thermal load is then implemented in the thermo-mechanical model to analyze the thermo-elastic behavior of the telescope. This is discussed in the next sections.


\section{Thermal Control and Thermal Model} \label{SS-TCTM}

Space observations offer many benefits such as the removal of all atmospheric optical effects. Unfortunately, these advantages come with a batch of drawbacks, one of the largest being ``thermo-mechanico-optical'' issues.
As the Sun--Earth distance varies naturally by 1.6\% along the year, the main thermal load is modulated by $\sim$3.2\% and the location of the abrupt transition created by the steep solar limb on the near-focus optical elements and on the CCD varies accordingly.

Additionally, the whole platform is further subject to thermal inputs from the Earth. They vary along the orbit because the platform maintains a fixed attitude with respect to the Sun and therefore not fixed with respect to the Earth, and because the terrestrial surface is not homogeneous (lands and seas, {\it etc.}). They also vary along the weeks and the year since the Earth climate has seasons and changing weather conditions. Finally, various degradations can affect the platform and the instrument, with consequences on their sensitivity to impinging thermal radiations. Consequently, a space telescope in low Earth orbit (LEO) is exposed to a thermal environment that evolves at many temporal scales.

Yet, if not controlled, such evolutions would turn deleterious for the wanted astrometric precision. Indeed, changes in the thermal input would convert into bending and other geometrical distortions, which themselves would have detrimental impacts on the optical performance (aberrations, optical distortion, magnification).

This is why the adopted strategy for SODISM has been to thermally stabilize all that can be controlled inside the telescope.
Nevertheless, some parts of the telescope (especially the door, and front window) cannot be regulated and a thermal model of the instrument is necessary to derive appropriate corrections and quantify the associated uncertainties

\subsection{The thermal control system (TCS)} \label{SS-TCS}

Thermal control is essential to guarantee the best possible performance and eventual success of the mission, but it is also needed to keep some subsystems safe, within required temperatures.
A thorough thermal strategy has therefore been implemented.
It consists of (i) an efficient thermal protection of the overall telescope {\it via} multilayer insulators (MLI), (ii) design and material choices that protect the exposed parts against solar illumination and further thermal issues, and (iii) the implementation of a complex thermal control system (TCS) that monitors many thermal sensors and drives several heaters.

The first two items pertain to passive thermal control: thermal insulation, radiators, surface finishes (coatings, paints, treatments, and second surface mirrors or SSM), and thermal washers which reduce the thermal coupling at platform interfaces.
The external blankets consist of 15 layers defined as follows: one external layer (space side, 25 $\mu$m Kapton), 13 intermediate layers (6 $\mu$m Mylar and Dacron) and 1 internal layer (internal side, 25 $\mu$m Kapton).
Degradation of the front side MLI is susceptible to occur as a result notably of solar irradiation.
The thermal radiators have been oriented in the direction of cold space (avoiding solar flux) in order to offer the best heat rejection capability.
Their external surfaces have been coated with the SG120FD white paint and SSM coating.
The mechanical parts on the optical path are black anodized according to ESA-PSS 01/703.
A lot of thermal shunt made of copper has been utilized at certain interfaces in order to provide a high conductive thermal coupling and a reduced structural stiffness.

\begin{figure}[h]
   \centerline{\hspace*{0.015\textwidth}
               \includegraphics[width=0.40\textwidth,clip=]{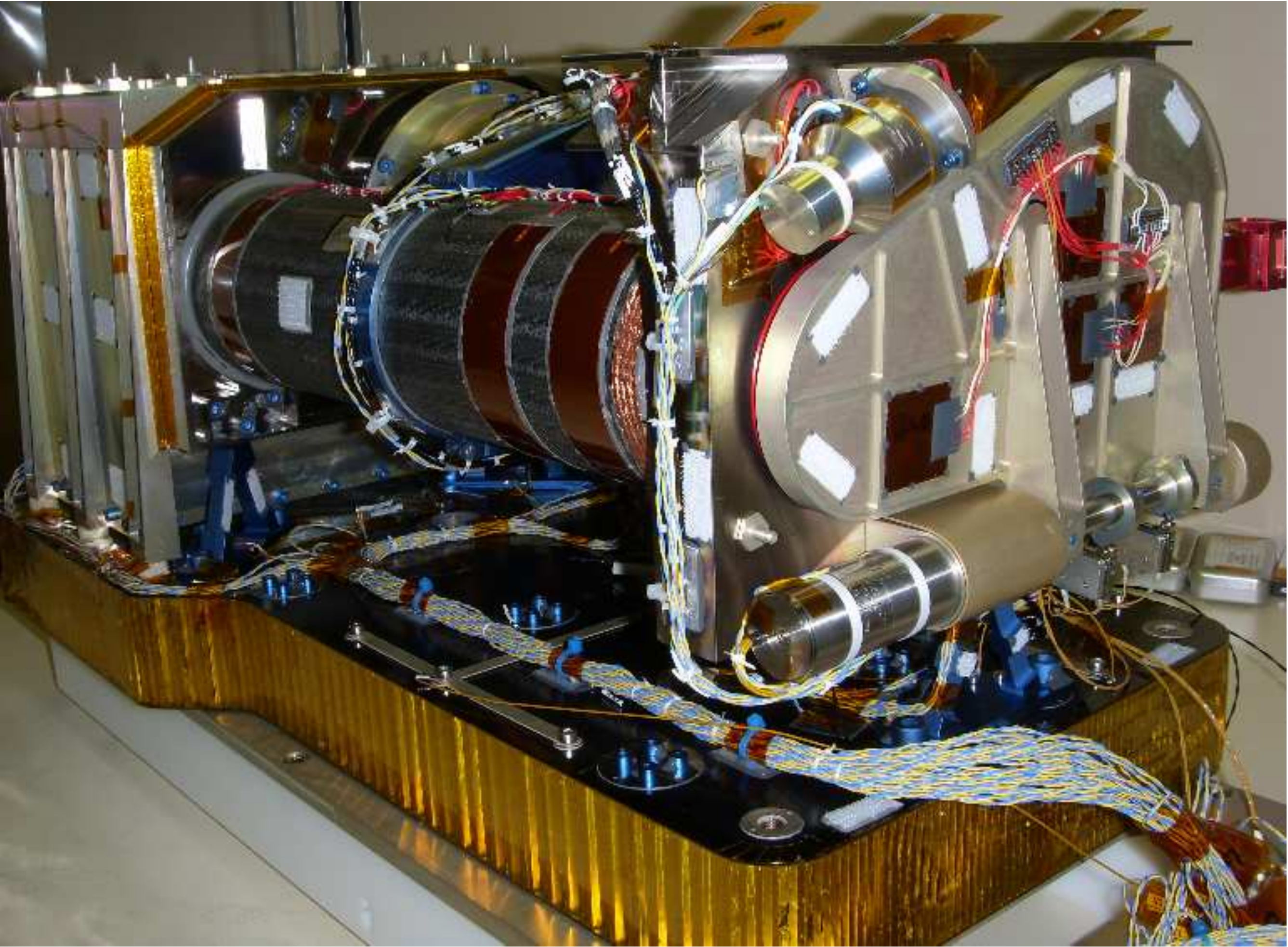}
               \hspace*{0.015\textwidth}
               \includegraphics[width=0.50\textwidth,clip=]{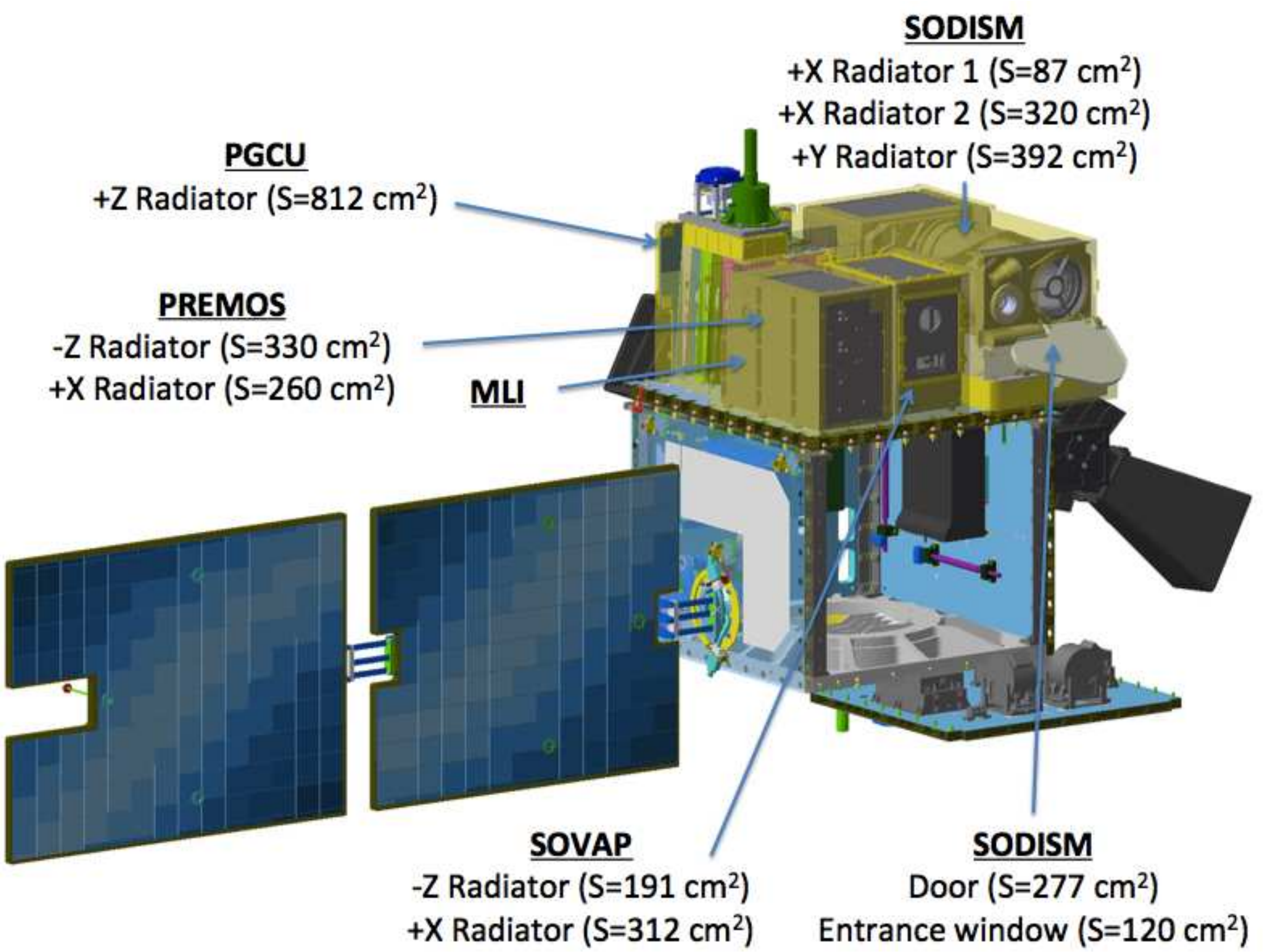}
              }
     \vspace{-0.35\textwidth}   
     \centerline{\small \bf     
      \hspace{0.0 \textwidth}  \color{black}{(a)}
      \hspace{0.39\textwidth}  \color{black}{(b)}
         \hfill}
     \vspace{0.31\textwidth}    
	\caption{SODISM during the integration is shown in Figure~\ref{figure_SODISM_integration}-a. Radiators of the spacecraft can be localized in Figure~\ref{figure_SODISM_integration}-b.
}
   \label{figure_SODISM_integration}
   \end{figure}

The PICARD payload includes several radiators. They are device with a large area which rejects the heat by radiating it to space.
It is desirable for their surface to have a high value of $\varepsilon$ (3 to 100\,$\mu m$ wavelength band) and a low value of $\alpha$ (0.2 to 3\,$\mu m$ wavelength band) in order to maximize their heat rejection and to minimize the solar inflow.
Thermal coatings are very efficient. But unfortunately, they degrade over time.
The locations of the radiators are shown in Figure~\ref{figure_SODISM_integration}-b.
Thermal control coatings surfaces, such as white paints, SSM (silvered or aluminized), Back Surface Mirrors (BSM), are used.
They have special radiation properties (Table~\ref{table_radiators_PICARD}).
Before launch, we made measurements to determine the value of $\alpha$ for the SG121FD white paint and found discrepant measurements.

Before SODISM was made operational, its stay-alive heaters prepared it for the unpowered phase (survival in cold conditions using thermo-switches and three heaters).
In non-operating mode, the in-orbit temperature range is foreseen to be [-35\,;\,+50]$^{o}$C. This avoids any degradation of the structure, electronics and optical parts.

The active temperature control system includes the hardware that is needed to heat and to measure the temperature of the controlled components of the telescope, including the structure, the mirrors, the mechanisms, {\it etc.}
But this system also implied a software development for the regulation algorithm.
The latter defines several areas that are located all over the structure and are independently controlled.
Each area consists of a set of heaters, mounted in the same electrical circuit, for which the heating power is adjusted based on the temperature measured by a sensor located in the vicinity.
Critical areas are controlled redundantly.
The temperature set point of each area is reprogrammable and was adjusted during the commissioning phase.
The thermal regulation is based on a Proportional Integral (PI) feedback.
The instrument stability strongly depends upon the precise definition of the parameters of these feedback loops.
The onboard electronic system calculates continuously the optimized electrical power needed to achieve the instrument regulation.

The heaters are of laminar type (copper resistive circuit encased in a Kapton film). They are bonded to the structure.
Critical heaters are redundant. This redundancy consists in a double resistive circuit, in order to minimize the heater surface on the structure.
The power density for each heater is less than 0.5\,W/cm$^{2}$.

The temperature sensors are used for the temperature regulation process, as well as for the telescope housekeeping, thereby enabling {\it a posteriori} thermo-mechanical modeling.
Analog temperature sensors (AD590MF/883B) have been used for high accuracy and high stability.
They have been calibrated so as to offer an accuracy of $\pm$0.1$^{o}$C.
There are 45 temperature sensors on the entire telescope and 3 for the electronics units.

Twenty heaters are used to thermally stabilize most regions of the instrument at 20$^{o}$C $\pm$1$^{o}$C.
In particular, the temperature of the interference filters is regulated at 20$^{o}$C $\pm$1$^{o}$C to avoid spectral shifts of their bandwidth.
But the CCD is thermally precisely regulated at -10$^{o}$C $\pm$0.2$^{o}$C to avoid any thermal expansion of its pixels, hence maintaining its geometry, and to limit the effects of dark signal and hot pixels.

The temperature of the front window cannot be regulated but the modeling predicts that it is kept by design between 0$^{o}$C and 40$^{o}$C while Sun-pointed.
It is indeed thermally insulated from the mechanical structure to avoid lensing effects that would be due to mechanical deformations or to temperature gradients caused by external constraints.
This is why radial and axial temperature gradients inside the window should be minimal.

The temperature of the prisms cannot be regulated either, but it is expected to stay between 10$^{o}$C and 35$^{o}$C.
However, the thermal stability of each prism is better than 0.05$^{o}$C per minute while the thermal gradient is less than 1$^{o}$C across one prism.
The temperature of each prism is measured with an accuracy of 0.3$^{o}$C.
Knowing the temperatures of the front optical elements (front window, green filter, prisms) permits to exploit metro-logically these subsystems.

A few problems remain to be solved, such as the accuracy of the temperature distribution and the effect of the degradation of the optics elements during the mission.
We foresee that space exposure will increase the absorbing properties of the optics and modify the temperature distribution.


\begin{table}[!h]
\caption{SODISM radiators.}
\label{table_radiators_PICARD}
\begin{tabular}{l l l l}
\hline
Denomination 							& Coating surface			& $\alpha_{BOL}$/$\alpha_{EOL}$ 				& $\varepsilon_{BOL}$/$\varepsilon_{EOL}$ \\
\hline
+X radiator 1						   	& Silvered SSM			& 0.14/0.18  								& 0.75/0.75					\\
+X radiator 2					  		& SG121FD white paint 		& 0.25/?\textsuperscript{$\star$} 				& 0.88/0.88					\\
+Y radiator (CCD)				     		& SG121FD white paint		& 0.25/?\textsuperscript{$\star$}				& 0.88/0.88					\\
Door							     		& Silvered SSM			& 0.14/0.18 								& 0.75/0.75					\\
Front window					     		& Optical coating 			& $\leq$0.21/?\textsuperscript{$\star$} 			& 0.81/0.81					\\
\hline
\end{tabular}
\quad \quad \quad \quad \quad \quad BOL: Begin Of Life, EOL: End Of Life (design hypothesis), ?\textsuperscript{$\star$} = Unknown ageing
\end{table}

\subsection{The thermal model and its coupling to the structural model} \label{SS-ThermalModel}

A thermal model allows estimating the temperature variations and the gradients experienced by the telescope in diverse situations.
It uses the dedicated Esarad and Esatan software packages from Alstom that operates by way of Monte Carlo ray tracing.
The thermal model hence computes the exchanges of energy between the surfaces of the geometrical model as well as the heat fluxes impinging on the instrument and the thermal radiation escaping from it to the cold deep space.
Obtained results are given in Table~\ref{T-SODISM temp simu}.

The parameters of the onboard TCS (Section \ref{SS-TCS}) were originally deduced from this model and from the thermal balance test.
They were subsequently adjusted during flight as satellite occultations by the Earth could be used as a temperature step response to estimate the instrument relaxation time which is an important parameter for the regulation.
The first ``eclipse season'' started in November 2010.

The output of the above thermal model can then be fed into the structural model (see section~\ref{SS-MechanicalModel}) in order to analyze the thermo-elastic behavior of the telescope.
This strategy was already used to assess the optical performance of SODISM during its thermal balance test.
Two typical cases of thermal configuration have been particularly studied: the Northern hemisphere winter and the Northern hemisphere summer loads.
These two different thermal cases, once coupled to the structural model, have allowed analyzing the thermo-elastic behavior of the telescope structure and its consequences on the optical path.

A NASTRAN software was also used to budget mechanical stability in the optical path (translations and rotations of the optics), and we could compute the sensitivity of the three main optical elements (M1, M2, and CCD) to a change of temperature of another element (Table~\ref{T-SODISM disp simu}).

\begin{table}[h!]
\caption{Temperatures and gradients provided by thermal analysis}
\label{T-SODISM temp simu}
\begin{tabular}{l l l}
\hline
Optical component and structure & Temperatures & Gradient\\
\hline
Front window					& -19$^{o}$C $<$ T $<$ 37$^{o}$C 		& 10$^{o}$C\\
Green Filter 					& 22$^{o}$C $<$ T $<$ 26$^{o}$C 		& 1.3$^{o}$C  \\
Prisms						& 22$^{o}$C $<$ T $<$ 24$^{o}$C 		& 1$^{o}$C \\
M2 Mirror 						& 29$^{o}$C $<$ T $<$ 34$^{o}$C 		& 1.18$^{o}$C \\
M1 Mirror 						& 23$^{o}$C $<$ T $<$ 25$^{o}$C 		& 0.22$^{o}$C\\
M3 Mirror 						& 18$^{o}$C $<$ T $<$ 22$^{o}$C 		& $<$ 0.2$^{o}$C\\
Photodiodes 					& 10$^{o}$C $<$ T $<$ 30$^{o}$C 		& $<$ 0.5$^{o}$C\\
Interference filter				& 18$^{o}$C $<$ T $<$ 24$^{o}$C 		& 0.1$^{o}$C\\
CCD							& -10$^{o}$C $\pm$0.2$^{o}$C 		& 2.5$^{o}$C\\
Mirror M1 plate					& 20$^{o}$C $\pm$5$^{o}$C 			& 3$^{o}$C\\
Mirror M2 plate					& 20$^{o}$C $\pm$5$^{o}$C 			& 7.9$^{o}$C\\
Filters wheels plate				& 20$^{o}$C $\pm$5$^{o}$C 			& 1.5$^{o}$C\\
CCD plate					& 20$^{o}$C $\pm$5$^{o}$C 			& 9.3$^{o}$C\\
Carbon Carbon tube				& 20$^{o}$C $\pm$5$^{o}$C 			& 1.1$^{o}$C\\
Carbon Carbon plate			& 20$^{o}$C $\pm$5$^{o}$C 			& 0.8$^{o}$C\\
\hline
\end{tabular}
\end{table}

\begin{table}[h!]
\caption{Displacement in the Z direction of M1, M2 and of the CCD for a +1$^{o}$C change in temperature on various elements of the telescope. This was obtained by thermo-mechanical analysis.}
\label{T-SODISM disp simu}
\begin{tabular}{l l l l l}
\hline
Heated element 		&	$\delta$T		&	dZ  M1 	&	dZ  M2		&	dZ CCD	\\
in the telescope		&				&	[nm]		&	[nm]			&	[nm]	\\
\hline						
Primary mirror M1		&	+1$^{o}$C	&	-150.00	&	-0.13			&	-0.18	\\
Secondary mirror M2	&	+1$^{o}$C	&	-0.03		&	25.30		&	-0.02	\\
CCD					&	+1$^{o}$C	&	0.00		&	0.00			&	-51.41\\
Carbon Carbon plate	&	+1$^{o}$C	&	-15.55	&	51.87		&	9.08	\\
Carbon Carbon tube		&	+1$^{o}$C	&	-31.89	&	35.81		&	-127.40	\\
Mirror M1 plate			&	+1$^{o}$C	&	-37.91	&	-96.88		&	-47.58	\\
Mirror M2 plate			&	+1$^{o}$C	&	58.69	&	81.18		&	30.96	\\
Filters wheels plate		&	+1$^{o}$C	&	-45.58	&	-41.76		&	37.34	\\
CCD support plate		&	+1$^{o}$C	&	-0.27		&	-0.17			&	87.42	\\
Piezoelectric support	&	+1$^{o}$C	&	-149.98	&	10.56		&	8.05	\\
\hline
\end{tabular}
\end{table}


\section{Design of the Management Unit and of the Flight Operations} \label{SS-Ops}

The PICARD payload is programmed by a main onboard electronics called the `PGCU' (``PICARD Gestion Charge Utile'' which stands for ``Payload Management Unit'').
The PGCU receives the tele-commands sent by the PICARD Payload Data Centers in Brussels and Toulouse \cite{Irbah2010,Pradels2012AcAau} and process them sequentially.
CCD images can be compressed onboard the PGCU in different ways before being transmitted.
The Langevin (LGV) algorithm \cite{Langevin2000SPIE,Said1996ITIP} has been implemented as part of the onboard software and is normally set at a rate of 16, although rates of 8 and 4 can be programmed too.
The LGV compression is nominally used for full-field full-resolution (FFFR) images, except for the defocused and the dark signal frames.
The lossless compression is the PLLC (PICARD lossless cruncher), which is CNES proprietary developed by E. Remetean.

During commissioning, the exposure duration for each channels were confirmed to be optimal at 1.2\,s for `215', 1.3\,s for `393', 1.0\,s for `535D', 6.9\,s for `535H', 0.97\,s for `607', and 1.1\,s for channel `782'. These values were used for quasi all ulterior image acquisitions, except for the two UV channels for which sensitivity degradation had to be compensated by successive increases of their exposure times.

Everyday operations consist in programming SODISM observations on the basis of a fixed daily template, the so-called {\it Routine} composed of a few observational programs:
\begin{itemize}
\item {\it The full frame full resolution synoptic monitoring:} In the frame of SODISM Routine, three daily full-field full-resolution (FFFR) images are produced in each of the six channels, except for the `393' one that delivers about eleven FFFR images per day instead of six.
These FFFR images are lossily (LGV) compressed onboard.
The primary role of the FFFR observations is to monitor the solar surface, and particularly the differential rotation and the various expressions of magnetic activity.
\item  {\it The macro-pixel (MP) program:} A second program involves also full-field images: the MP series generates `535H' images once per minute. MP images are binned onboard into 256$\times$256 frames. The primary application of the MP sequence is on-disc helioseismology.
\item {\it The HL narrow annulus program:} The next program of the Routine supports helioseismology too. One `535H' MP acquisition out of two is not only binned into an MP image, but also cropped into a narrow annulus image in order to spare the telemetry budget. These data are losslessly compressed and they are dubbed "HL". The annulus used to be 22-pixels thick until March 2012, when it augmented up to 31 pixels henceforth.
\item {\it The DL wide annulus program:} The last dataset produced by the Routine corresponds to the so-called DL annuli images. They are carried out in all channels (contrarily to HL that occurs in `535H' only). Their thickness has been 40 pixels all along, and their frequency has been 22 observations per channel and per day.
\item {\it The HMP program:} 1-mn sampling, interrupted by the other monitoring.
\item {\it The HL program:} 2-mn, with no interruptions but the other modes.
\end{itemize}
Some calibration data are also acquired in the context of the Routine: dark signal and defocused images. The latter employ the defocusing lens available on the second filter wheel. The duration of integration of the former has been cycling after September 2012. It must finally be mentioned that the Routine program adapts to the ``eclipse season'', {\it i.e.} from mid November to early February. In particular, FFFR images of the Sun partly absorbed by the Earth atmosphere are recorded; this is the so-called MAB mode (Absorption mode).

SODISM special operations belong to two types. The first type of manual procedures corresponds to the commissioning phase (June-September 2010). The second category gathers all other calibration campaigns. An exhaustive listing of all special operations would fall beyond the scope of the present paper.
Yet, it is worthwhile mentioning those which have a significant calibration role:
\begin{enumerate}
\item {\it Spacecraft rolling (also known as `MDO'):} During such campaigns, the whole mission and its payload revolve around the PICARD-Sun axis by steps of 30$^o$. Such campaigns occurred several times in different channels. They aim at estimating the asphericity -- and particularly the oblateness -- of the solar disc. Besides, they contribute to the calibration of the field of distortion and to large scale flat-fielding. As compared to the original plan, the MDO procedure has been improved to permit revolving PICARD twice, instead of only once, or alternatively, to double the repetend duration (roll maneuver + data acquisition) from one up to two orbits. In addition, one of the MDO campaigns produced FFFR images instead of the normally planned annuli images.
\item {\it Stellar pointing (aka MES):} In addition to its corner images (which provide a way to investigate possible evolutions of the telescope magnification), PICARD is able to overturn and observe stars in the night sky. Doublets of stars, separated by {\it circa} a solar angular diameter, were selected before launch and hard-coded in the operational scheme. This was meant to offer an absolute calibration of SODISM magnification, but it provides other benefits to SODISM, and to the rest of the payload too.
\item {\it Scan of the exposure time:} In few occasions, the exposure time was scanned from it minimum value (0.5\,s) up to saturation. This is important to discriminate the effects of the stabilization systems. Additionally, it allows studying tradeoffs and the mixed effects of a larger SNR, higher cosmic ray hits (CRHs) density, and LGV compression rates.
\item {\it Burst mode:} Normally, routine operations bring forth three FFFR images per day (eleven in the case of the `393' channel). This cadence is insufficient to catch flares or to precisely track the motion and brightness variations of sustained features. High cadence sequences were  run (between May 17th and May 20th, 2012) at the maximal cadence of one image per minute, until all onboard buffers were filled up, which allowed taking data for 4 to 8 hours depending on the channels.
\item {\it Internal off-pointing:} The piezo actuators have been  commanded to impose rotational scans of the M1 mirror, resulting in mild amplitude translations of the Sun in the FOV. Such sequences enable the reconstruction of the instrumental flatfield (combining filters and CCD flat-fields): see section~\ref{SS-InternalPointingInterest}.
\item {\it Piston mode:} The piezos can also be collectively actuated in order to impose a translatory motion of the M1 mirror along its Z optical axis (instead of a rotation around the X and Y axis as mentioned in the previous bullet). This results in an exploration of the focusing.
\item {\it Spacecraft off-pointing:} Although they require special preparation at CNES, large off-points (enabled by platform maneuvers) allow exceeding the limited amplitude of M1 rotations. This is useful in order to {\it e.g.} position the corner images in the center of the FOV. This special operation occurred on March 11th, 2012.
\item {\it Bake-outs:} This special procedure heats the CCD temperature up to $\sim$$20^o$C. It does not allow removing possible contaminants that have probably been polymerized in the meanwhile, but it does allow passivating partly the hot pixels. The ones of the imaging zone (IZ) are only reduced in amplitude, but more importantly, the hot pixels of the memory zone (MZ) too; this has the beneficial effect of lowering the main component of the dark signal and associated noise. Bake-outs were performed in June 2011 (2 days) and in June 2012 (3 days).
\end{enumerate}


\section{SODISM in space and preliminary observational results} \label{SS-InSpace}

The PICARD spacecraft is on a Sun synchronous orbit at an altitude of $\sim$735 km, allowing to observe the Sun non-stop, except for short periods during a November-January `eclipse season' when the Earth comes within SODISM line-of-sight once per orbit.
The spacecraft AOCS is continuously operated to maintain the solar rotation axis in a constant direction, {\it i.e.} ordinarily the solar poles are in a direction parallel to the vertical axis of SODISM images.

\begin{figure}[h]
	\centerline{\includegraphics[width=1.0\textwidth]{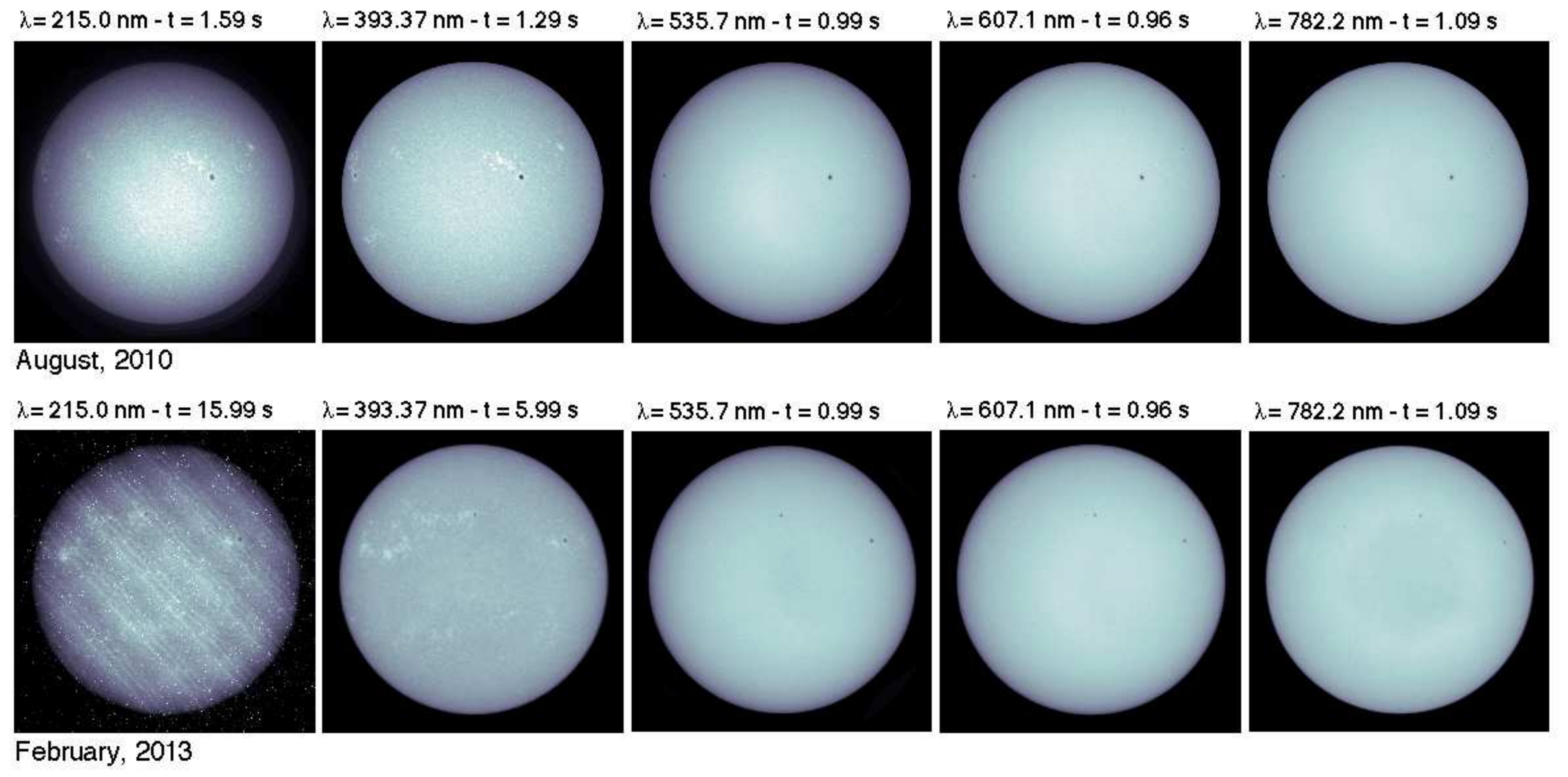}}
	\caption{Solar images in the different SODISM channels (Level 1 data products). The upper row corresponds to August 2010, near beginning of life. The lower row corresponds to February 2013.}
	\label{Fig_SODISM_Images_Mission_Picard}
\end{figure}

\begin{figure}[h]
	\centerline{\includegraphics[width=1.2\textwidth]{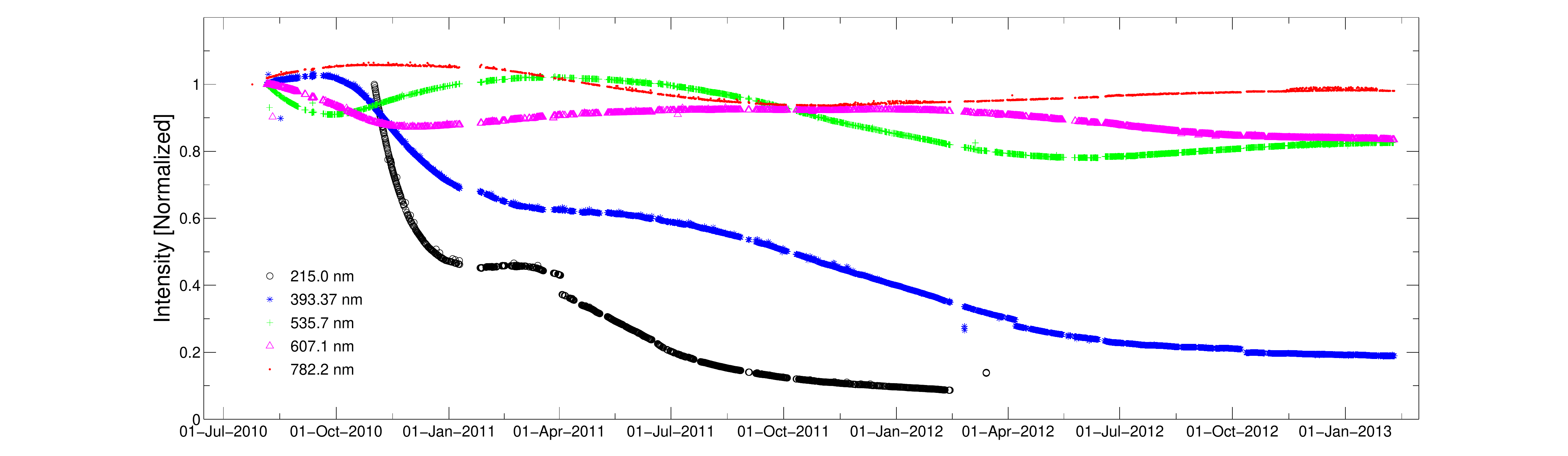}}
	\caption{Normalized time series of the integrated intensity since the beginning of the mission. Black is at `215', blue at `393', green at `535', pink at `607', and red at `782'.}
	\label{Fig_SODISM_Images_Intensites}
\end{figure}

SODISM door was opened in July 2010 (after some faltering). Its two filter wheels and its shutter mechanisms operate 24-7-365 flawlessly since then, almost every minute.
As to its flight software, no substantial errors were encountered; the commanding and the onboard data handling perform as they were intended to.
The Level\,0 (L0) data products are correctly generated every day at the PICARD Payload Data Center situated at BUSOC in Brussels \cite{Irbah2010}.
SODISM has recorded more than one million solar images since the beginning of the mission in June 2010 and is still operating in 2013.
The upper part of Figure~\ref{Fig_SODISM_Images_Mission_Picard} shows a sample of Level\,1 solar images at the five wavelengths recorded in August 2010, while the lower part represents solar images recorded in February 2013. 
All have been corrected for dark current and flatfield (using the method proposed by Kuhn, Lin, and Loranz \citeyear{Kuhn1991PASP}).
We observe in Figure~\ref{Fig_SODISM_Images_Mission_Picard} that `215' and `393' images exhibit in 2013 a flatter Center to Limb Variation (CLV) than in 2010.
This is due to the expectable loss of sensitivity in these ultraviolet channels.
It is interesting to monitor the integrated intensity in SODISM images at all wavelengths.
This has been computed since the beginning of the mission and displayed in Figure~\ref{Fig_SODISM_Images_Intensites}, where UV channels are seen to experience degradation. 
By early 2013, the `215' channel has lost more than 90$\%$, and `393' about 80$\%$.
The degradation is probably induced by the polymerization of contaminants on the front window and/or on the other optical elements under solar UV exposure.
Note that the integrated intensity for the visible and near infrared channels presents a temporal oscillation but remains relatively constant.
We are currently investigating why the degradation is not a decreasing function for all channels.

\begin{figure}[h]
	\centerline{\includegraphics[width=1.1\textwidth]{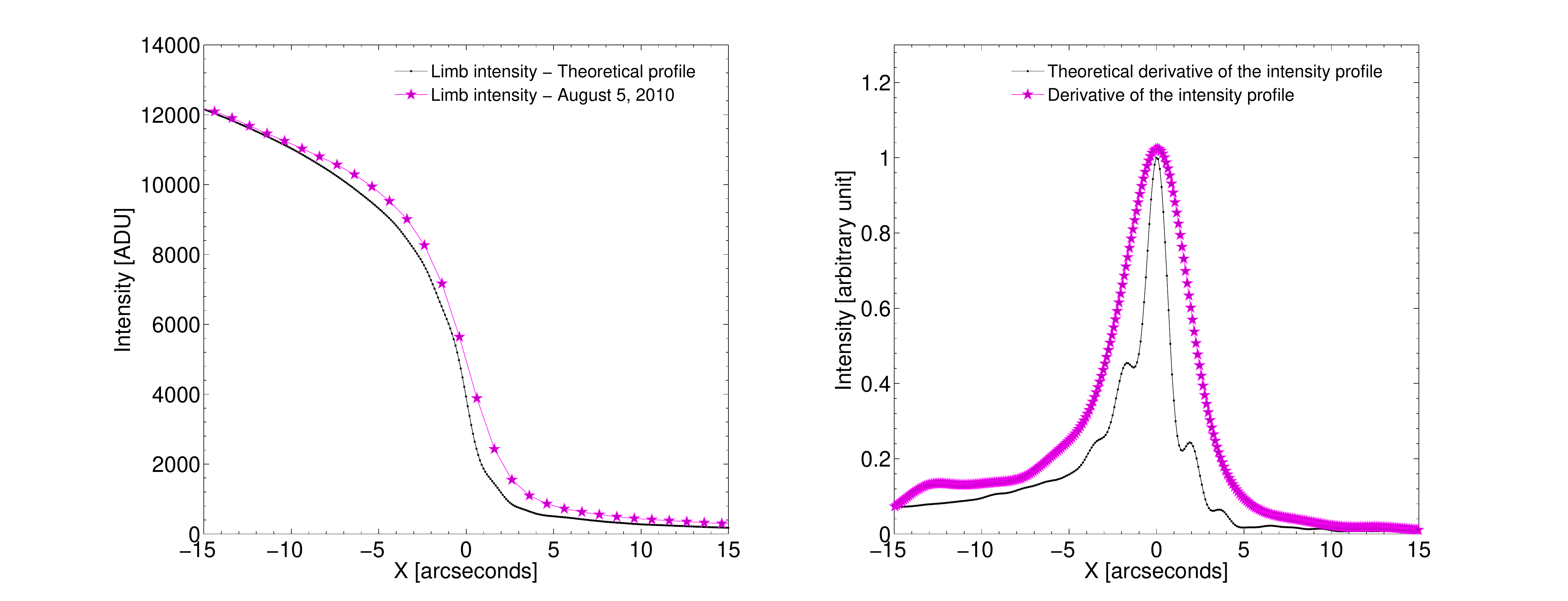}}
	\caption{Left panel: limb darkening function (LDF) at 607.1\,nm, and right panel: first derivative of the LDF at 607.1\,nm. 
                         The Hestroffer \& Magnan (1998) parametric model of the LDF, convolved with the theoretical PSF of SODISM is represented by continuous black curves.
                          August 2010 data are overplotted in pink.}
	\label{Fig_SODISM_Images_Limb_607nm}
\end{figure}

\begin{figure}[h]
	\centerline{\includegraphics[width=1.2\textwidth]{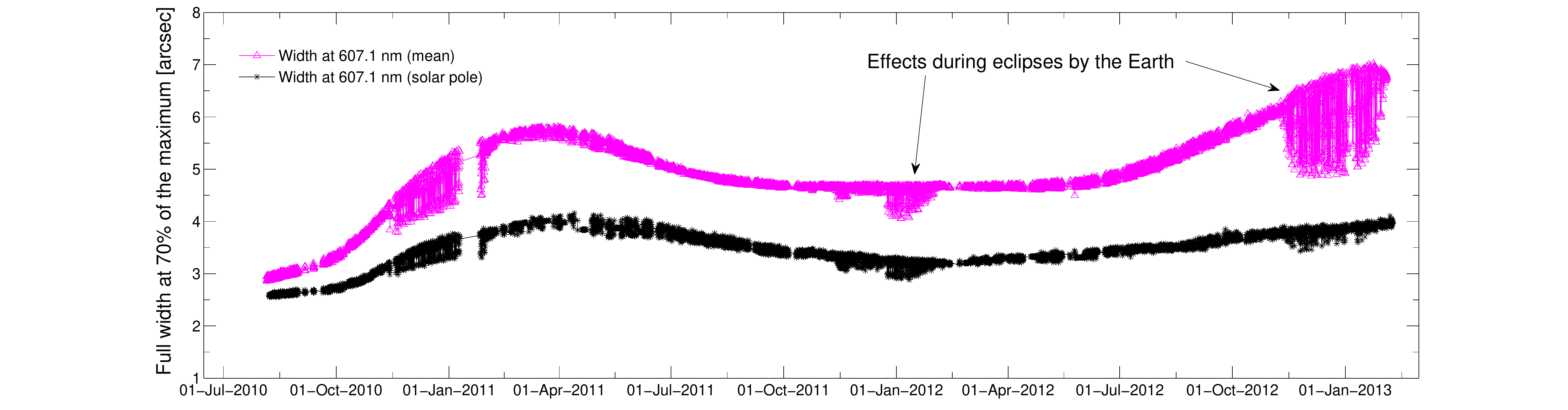}}
	\caption{Evolution of the limb width at 607.1\,nm since the beginning of the mission. The spread is measured as the full width at 70$\%$ of the maximum of the first derivative. The black curve is at the solar poles; the pink curve is the average along the circumference.}
	\label{Fig_SODISM_Images_Evolution_Limb_607nm}
\end{figure}

We also study the quality of the first SODISM images to see whether the solar diameter measurement can be achieved with those data.
We consider below solar images recorded at 607.1\,nm. We first extract radial cuts in the image, which show the solar limb intensity darkening as seen by the instrument.
We compare them with an empirical parametric model of the limb darkening function (LDF) \cite{Hestroffer1998AA} convolved with the nominal PSF of the instrument. 
The PSF is obtained with an optical model of the instrument designed with the ZEMAX software \cite{Irbah2010}.
The left side of Figure~\ref{Fig_SODISM_Images_Limb_607nm} compares the convolved LDF at 607.1\,nm with the LDF that is actually observed. 
We next compute the first derivative of those two LDFs to better evaluate their spread (right side of Figure~\ref{Fig_SODISM_Images_Limb_607nm}). 
The observational curve gives an indication of image quality.
We notice that the solar limb recorded with SODISM is wider than the spread expected from the model. 
The slope of the observed intensity profile --\,defined as the Full Width at Half (or 70$\%$) Maximum (FWHM) of its first derivative\,-- is wider than expected by $\sim$3\,arcsec. 
This is probably due to a misalignment of the optical elements (see Table~\ref{table_limb_width_Effect}), combined with thermo-optical effects, which can also blur the image. 
In itself, this does not disqualify the scientific objective since it is equivalent to having a telescope with a smaller aperture, giving the same PSF as observed.
But it would have to be time invariant, {\it i.e.} constant during the entire mission.
However, the limb width evolves with time: it displays a modulation in phase with the orbit
due to the Earth atmospheric radiations, which affect the observations \cite{Irbah2012SPIE},
as well as a long-term trend toward worsening (Figure \ref{Fig_SODISM_Images_Evolution_Limb_607nm}).
Additionally, the solar limb at the equator is more spread out than the limb at the pole.

The temporal variations shown in Figure~\ref{Fig_SODISM_Images_Evolution_Limb_607nm}, do not have a random behavior.
They are susceptible to be modeled, and it is actually our challenge to develop the model that will allow correcting the data for the instrumental changes.

The Venus transit has been observed with PICARD On June 5-6, 2012.
At this occasion, special observations were planned and they appear already very useful to the PSF calibration and modeling purpose.

We look now at the impact of the mentioned image quality issue on the PICARD helioseismology program. 
From the first days of in-flight PICARD operations, we continuously produce $l$-$\nu$ diagrams from the set of Macro-Pixel images.
The ridges are well defined up to $l$=300. 
With long enough time series and a duty cycle above 70\%, this allows to conduct the  medium-$l$ program \cite{Corbard2008AN}, {\it i.e.} to fit mode parameters  and perform inversion for the internal rotation profile \cite{Corbard2013}. 
Therein, 769 oscillation modes with spherical harmonics up to $l$=99 are fitted and frequency splittings inverted showing the differential rotation rate throughout the convection zone and the steep gradient of the tachocline.
The image blur has no negative impact on this program which uses binned images. 
It actually reduces the spatial aliasing.

\begin{figure}[h]
	\centerline{\includegraphics[width=0.75\textwidth]{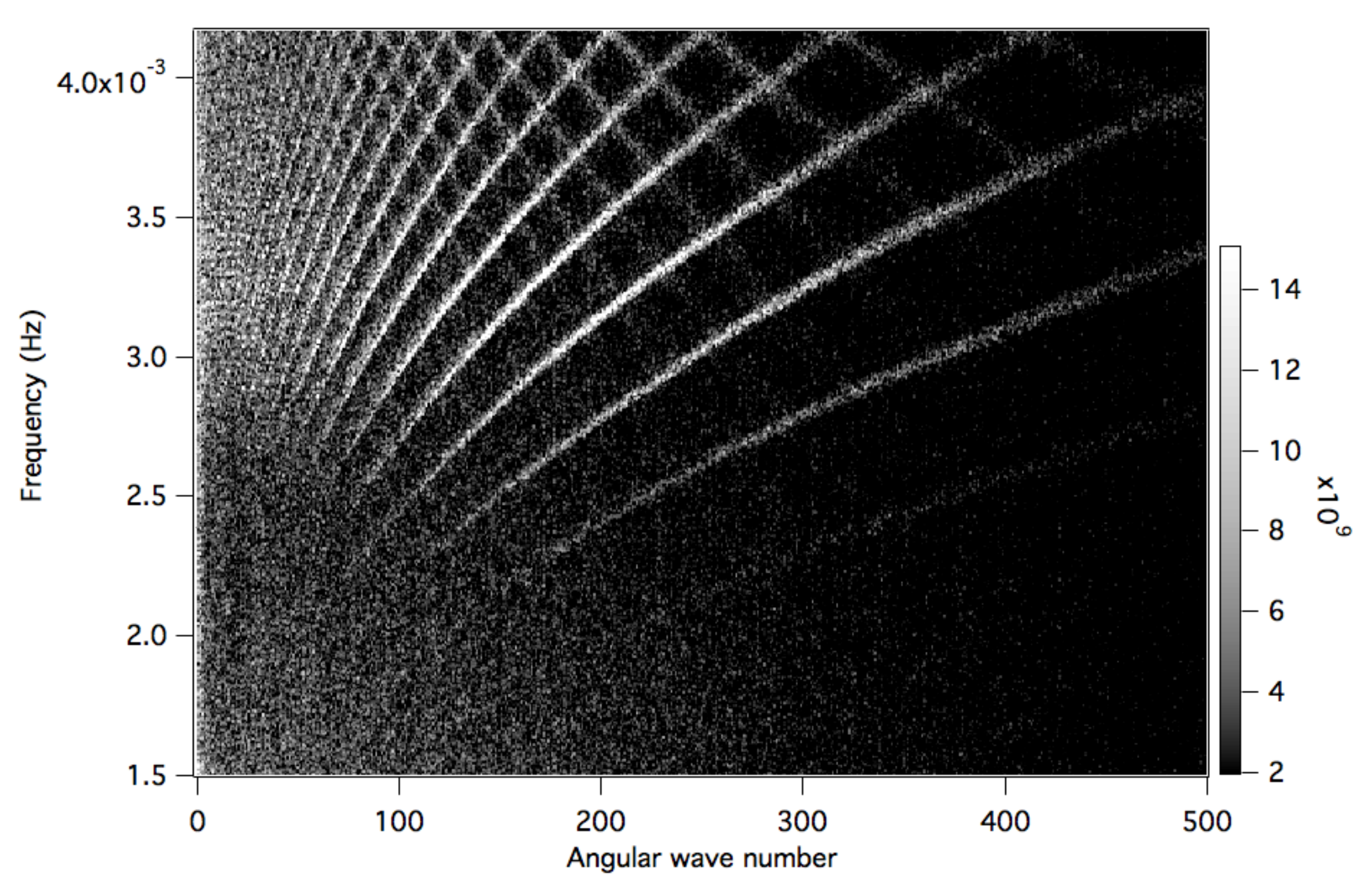}}
	\caption{$l$-$\nu$ diagram computed from three days of SODISM observations. 
                         In April 2011, 3 consecutive days were dedicated to helioseismic measurements at 535.7\,nm without interruption.}
	\label{Fig_SODISM_helio_limb}
\end{figure}

But, for the program of helioseismology at the solar limb, it is important to show whether or not the instrumental defocussing prevents the detection of p modes at the extreme limb, where we know they can be amplified \cite{Toutain1999}.
To answer this question, SODISM data recorded in April 2011 at 535.7\,nm have been analyzed. 
They consist in a ring of 22-pixel width centered on the solar edge and recorded every 120 seconds during 3 consecutive days. 
The analyzed data series has a duty cycle of 95$\%$. 
Each ring image is divided into several sectors (here 600) over which the intensity is computed on a smaller ring of 10 pixels centered on the solar image. 
All the data set is then summarized in a matrix of 600 $\times$ 2160. 
A 2-D Fourier Transform is made on the obtained matrix leading to the result shown in Figure~\ref{Fig_SODISM_helio_limb}. 
This clearly shows that the p modes are also detected in SODISM limb images recorded every two minute.

Two other result papers, related to the astrometry program, are in preparation: a study of the June 2012 Venus transit for solar diameter determination, and another one on the solar oblateness.
In addition, SODISM data can provide an estimate of the parameters of the differential rotation and meridional flow as a function of time and wavelength.
A previous work allows detecting efficiently the features in solar images \cite{Djafer2012}. This method is now used to analyze the SODISM data together with the solar images recorded at the same date with the PICARD ground-based telescope \cite{Meftah2012SPIE}. Such approach will be the subject of future papers.

\section{Concluding remarks} \label{S-Conclusion}

At the time of this instrumental paper writing (2012 and early 2013), SODISM has already been performing space observations for more than two years. The present redaction comes thus late as compared to the usual timeline for equivalent programs. However, this delay gave the possibility to provide in these pages a large view of the design, development, flight operations, and some observations.
SODISM discloses a very good instrumental health.
Unfortunately, the quality of raw images reveals an optical aberration that blurs the images in an evolving way. Its analysis and partial correction will be the subject of other publications.
The degradation of space instruments is complex, their causes and mechanisms are, in many instances, difficult to understand, since they can be the result of a combination of independent degradation effects.
SODISM does not escape to this rule, but there are important lessons to learn from it concerning in-orbit degradation of solar instruments.

Nevertheless, SODISM provides a distinctive opportunity to investigate solar oscillations from intensity variations observed at the limb and on the disc.
This investigation and the comparison with studies using the velocity signal is important because it provides useful information on the instrumental
and solar background noises. This is a way
to improve our understanding of the physics of solar oscillations  and their interaction with convection and the solar atmosphere and this can contribute
to more reliable interpretation of helioseismic results.

We recognize also another scientific objective achievable by the instrument: the asphericity at several wavelengths.
The magnitude of the solar oblateness is an important parameter to constrain models of the solar interior. 
However, this is a difficult quantity to measure due to its very small value (the difference in equator to pole radius is less than 10$^{-3}$\,arcsec). 
The distortion of SODISM solar images may be two orders of magnitude larger.
Yet, the `MDO' distortion mode (see Sect.~\ref{SS-Ops}) allows determining and removing the effect of distortion in images. 
In this mode the satellite rotates around its axis directed to the Sun with 12 positions separated by 30$^o$ and the polar and equatorial axes of the Sun takes alternatively the 12 possible orientations on the CCD image. 
To improve the oblateness determination, each satellite position is kept during 2 orbits since July 2011.
A distortion mode is made in less than 2 days and the drift of instrument parameters may be neglected. 

Another achievable objective is to provide a determination of the solar diameter from SODISM, at least from the Venus transit data.
They indeed offer unique opportunities to determine accurately the solar diameter from the duration of the transit. 
The relative position of Venus and the Sun are known with very high accuracy from ephemerids provided by the IMCCE, which allows to consider a determination of the solar diameter with low uncertainty.
These data provides also the possibility to estimate the centre to limb variation of the solar flux. 
SODISM took full images at four wavelengths (393.37, 535.7, 607.1, and 782.2\,nm, with some preference for 607.1\,nm) during the ingress and egress of Venus.

Finally, various other investigations can be performed such as the evolutions of the supergranular pattern, differential rotation in all spectral channels, chromospheric studies, to mention but a few.

The SODISM archive presents a great scientific value and is now accessible to the scientific community on simple request to the Principal Investigator of the PICARD mission, Dr Alain Hauchecorne, until it is made available online.


\begin{acks}
PICARD is a mission supported by the CNES, the CNRS/INSU, the Belgian Space Policy (BELSPO), the Swiss Space Office (SSO), and the European Space Agency (ESA).
The SODISM instrument has been built by CNRS - LATMOS.
We thank CNES, and CNRS for their support as well as all participants having devoted their expertise to this project. We wish to express our gratitude to the members of the calibration facility station of IAS (France) who participated to the preparation of this mission, and the help of the LESIA laboratory (France).
Numerous individuals have been involved in this project (Claude Leroy, Emmanuel Ducourt, Ludovik Bautista, ...).
\end{acks}


\bibliographystyle{spr-mp-sola}
\bibliography{SODISM_meftah_et_al_JFH}

\end{article}

\end{document}